# Understanding Social-Force Model in Psychological Principles of Collective Behavior

**Peng N. Wang**

***Abstract—***

To well understand crowd behavior, microscopic models have been developed in recent decades, in which an individual's behavioral/psychological status can be modeled and simulated. A well-known model is the social-force model innovated by physicists (Helbing and Molnar, 1995; Helbing, Farkas and Vicsek, 2000; Helbing et al., 2002). This model has been widely accepted and mainly used in simulation of crowd evacuation in the past decade. A problem, however, is that the testing results of the model were not explained in consistency with the psychological findings, resulting in misunderstanding of the model by psychologists. This paper will bridge the gap between psychological studies and physical explanation about this model. We reinterpret this physics-based model from a psychological perspective, clarifying that the model is consistent with psychological studies on stress, including time-related stress and interpersonal stress. Based on the conception of stress, we renew the model at both micro-and-macro level, referring to agent-based simulation in a microscopic sense and fluid-based analysis in a macroscopic sense. The cognition and behavior of individual agents are modeled as response to environmental stimuli. Existing simulation results such as faster-is-slower effect will be reinterpreted by Yerkes–Dodson law, and herding effect, social groups as well as oscillation phenomenon are especially discussed for pedestrian crowd. In brief the social-force model exhibits a bridge between the physics laws and psychological principles regarding crowd motion, and this paper will renew and reinterpret the model on the foundations of psychological studies.

## I. Introduction

Collective behavior of many individuals is a typical research subject in complex systems, and it is of great theoretical and practical interest. One of the major approaches to study such a system is describing individual behavior by a mathematical model and using computer programs to simulate how collective behavior on a macroscopic scale emerges from individual interactions. This approach is usually called agent-based modeling (ABM) or agent-based simulation (ABS), and it has been widely used in many diverse disciplines, including sociology, economics, molecular biology and so forth. In statistic physics this approach is also named by many-particle simulation (Helbing, 2001), where each individual is represented by a particle, and their collective behavior is studied by simulation of many particles.

A well-known model to simulate complex dynamics of pedestrian crowd is called social-force model, and it was initally introduced by physical scientists at the end of last century (Helbing and Molnar, 1995; Helbing, Farkas and Vicsek, 2000; Helbing et al., 2002). Their innovations are mostly from physics, especially from molecular physics and fluid physics. Very interestingly, certain psychological factors are abstracted in model parameters such as "desired velocity" and "social force." To some extent, these concepts are not physical entities because they describe people's opinions or certain characteristics of human mind. Thus, the model is not typically within the scope of physics study, but in an interdisciplinary domain, and an important question is whether such innovation is consistent with psychological theories or findings on human.

---

Peng N. Wang previously studied in the Department of Electrical and Computer Engineering, University of Connecticut, Storrs. Parts of the research work were conducted in his graduate program in University of Connecticut. The research was partly supported by NSF Grant # CMMI- 1000495. If this manuscript is helpful to your research, please consider citing it as
P.N. Wang, "Understanding Social Force Model in Psychological Principles of Collective Behavior," arXiv:1605.05146, Master Thesis, 2016.

On one side, the social-force model has been widely used in simulation of crowd evacuation in the past two decades. Many leading simulators use or extend the social-force model as the pedestrian model such as FDS+Evac (Korhonen et. al., 2008, Korhonen, 2017), PedSim, MassMotion (Oasys, 2018) and so forth. On the other side, some controversial issues have been long raised for the model. For example, the anisotropic social-force disobeys the Newton 3rd law and this is questionable in physics because pedestrian motion are definitely governed by Newton Laws. The controversial issue seems difficult to be explained by the model itself, especially from the perspective of physics. Thus, in this contribution we will try to provide a new perspective to understand the model, and the key idea is borrowed from psychological study of stress.

Actually stress was initially not a psychological term, but a physical measure, which expresses the internal forces in a body between its particles. This term was borrowed from physics to psychological circles since Selye's 1930s experiments, where the term was referred to the interaction between the environment and the living bodies, emphasizing the role of the individual's appraisal of situations in shaping their responses. Now we would like to use this psychological concept to understand the social-force model. Based on the conception of stress the model will be renewed in several aspects, referring to both of self-driving force and social force, and a new concept of desired interpersonal distance is critically introduced, by which we will explain the model agrees with Newton 3rd Law for pedestrian motion. In order to justify our explanation about stress, we will reiterates the faster-is-slower phenomenon by Yerkes-Dodson law, and the faster-is-slower effect actually shows a doorway scenario where excess of stress impairs human performance in the collective sense. The detailed discussion will be presented in Section 3.

With the new conception on stress the social force model will be discussed in both microscopic and macroscopic sense, and a micro-and-macro linkage will be established.

At the microscopic level the model describes an individual behavior, from which crowd phenomenon will be simulated in the collective sense. In particular we mainly study how an individual interacts with the surrounding individuals, and herding and grouping effects will be discussed in detail in Section 4. Interestingly enough, opinion dynamics will be integrated into the social force model to characterize "herding effect", and a new formula of group social force is presented to characterize group dynamics in pedestrian motion. The force is an extension of the traditional social-force, and it characterizes the social relationship among individuals. In order to offset unexpected oscillation effect the model is further investigated by ordinary differential equation (ODE) in Section 5.

At the macroscopic level we will aggregate the Newtonian-like individual model into the fluid-like crowd model in Section 6. The micro-and-macro link between the social-force model and fluid model is explored for high-density crowd. The fluid-based analysis provides another perspective by using partial differential equation (PDE), and it continues with the key concept of stress, and it enables an energy viewpoint to understand crowd behavior: the psychological drive arises as a potential form of energy, and it will be transformed to energy in kinetic form or static form, resulting in motion or compression of the crowd. An energy balance equation is derived in Section 7, and the faster-is-slower effect is revisited from such energy viewpoint. By using the energy equation we explain why stampede often happened in stadium and two major reasons include "heating effect" and "downstairs tragedies."

In brief the collective behavior of crowd has long been investigated in psychology and social science, and there were a great many valuable theories and findings. As the leading model to study complex dynamics of pedestrian crowd, the social force model is to be inherently consistent with both physics laws and psychological principles, and this article will try to establish a bridge between two research areas, and cast a new light on understanding of the social-force model from a new perspective.

## II. About the Social-Force Model

Physics is the analysis of nature, conducted to understand how the universe behaves, while psychology is the study of the human mind and behaviors, exploring how human thinks and behaves. In the past decade a pedestrian model involves both of these two subjects in the framework of Newtonian dynamics, and the model is called the social-force model (Helbing and Molnar, 1995; Helbing, Farkas, and Vicsek, 2000; Helbing et al., 2002).

The social-force model presents psychological forces that drive pedestrians to move as well as keep a proper distance with others. In this model an individual's motion is motivated by a self-driving force $\boldsymbol{f}_i^{drv}$ and resistances come from surrounding individuals and facilities (e.g., walls). Especially, the model describes the social-psychological tendency of two individuals to keep proper interpersonal distance (as called the social-force) in collective motion, and if people have physical contact with each other, physical forces are also taken into account. Let $\boldsymbol{f}_{ij}$ denote the interaction from individual $j$ to individual $i$, and $\boldsymbol{f}_{iw}$ denote the force from walls or other facilities to individual $i$. The change of the instantaneous velocity $\boldsymbol{v}_i(t)$ of individual $i$ is given by the Newton Second Law:

$$m_i \frac{d\boldsymbol{v}_i(t)}{dt} = \boldsymbol{f}_i^{drv} + \sum_{j(\neq i)} \boldsymbol{f}_{ij} + \sum_{w} \boldsymbol{f}_{iw} + \boldsymbol{\xi}_i$$

$$m_i \frac{d\boldsymbol{v}_i(t)}{dt} = \boldsymbol{f}_i^{drv} + \sum_{j(\neq i)} (\boldsymbol{f}_{ij}^{soc} + \boldsymbol{f}_{ij}^{phy}) + \sum_{w} (\boldsymbol{f}_{iw}^{soc} + \boldsymbol{f}_{iw}^{phy}) + \boldsymbol{\xi}_i \tag{2.1}$$

where $m_i$ is the mass of individual $i$, and the self-driving force $\boldsymbol{f}_i^{drv}$ is specified by

$$\boldsymbol{f}_i^{drv} = m_i \frac{\boldsymbol{v}_i^0(t) - \boldsymbol{v}_i(t)}{\tau_i} , \tag{2.2}$$

This force describes an individual tries to move with a desired velocity $\boldsymbol{v}_i^0(t)$ and expects to adapt the actual velocity $\boldsymbol{v}_i(t)$ to the desired velocity $\boldsymbol{v}_i^0(t)$ with a characteristic time $\tau_i$. In particular, the desired velocity $\boldsymbol{v}_i^0(t)$ is the target velocity existing in one's mind while the actual velocity $\boldsymbol{v}_i(t)$ characterizes the physical speed and direction being achieved in the reality. The gap of $\boldsymbol{v}_i^0(t)$ and $\boldsymbol{v}_i(t)$ implies the difference between the human subjective wish and realistic situation, and it is scaled by a time parameter $\tau_i$ to generate the self-driving force. This force motivates one to either accelerate or decelerate, making the realistic velocity $\boldsymbol{v}_i(t)$ approaching towards the desired velocity $\boldsymbol{v}_i^0(t)$. This mathematical description of the self-driving force could be dated back to the Payne-Whitham traffic flow model (Payne, 1971; Whitham, 1974). Sometimes $\boldsymbol{v}_i^0(t)$ is rewritten as $\boldsymbol{v}_i^0(t) = v_i^0(t)\boldsymbol{e}_i^0(t)$, where $v_i^0(t)$ is the desired moving speed and $\boldsymbol{e}_i^0(t)$ is the desired moving direction. In a similar manner, we also have $\boldsymbol{v}_i(t) = v_i(t)\boldsymbol{e}_i(t)$ where $v_i(t)$ and $\boldsymbol{e}_i(t)$ represent the physical moving speed and direction, respectively.

The interaction force of pedestrians consists of the social-force $\boldsymbol{f}_{ij}^{soc}$ and physical interaction $\boldsymbol{f}_{ij}^{phy}$. i.e., $\boldsymbol{f}_{ij} = \boldsymbol{f}_{ij}^{soc} + \boldsymbol{f}_{ij}^{phy}$ . The social-force $\boldsymbol{f}_{ij}^{soc}$ characterizes the social-psychological tendency of two pedestrians to stay away from each other, and it is given by

$$\boldsymbol{f}_{ij}^{soc} = A_i \exp\left[\frac{(r_{ij} - d_{ij})}{B_i}\right] \boldsymbol{n}_{ij} \quad \text{or} \quad \boldsymbol{f}_{ij}^{soc} = \left(\lambda_i + (1-\lambda_i)\frac{1+\cos\varphi_{ij}}{2}\right) A_i \exp\left[\frac{(r_{ij} - d_{ij})}{B_i}\right] \boldsymbol{n}_{ij} \tag{2.3}$$

where $\boldsymbol{A}_i$ and $\boldsymbol{B}_i$ are positive constants, which affect the strength and effective range about how two pedestrians are repulsive to each other. The distance of pedestrians $i$ and $j$ is denoted by $d_{ij}$ and the sum of their radii is given by $r_{ij}$ . $\boldsymbol{n}_{ij}$ is the normalized vector which points from pedestrian $j$ to $i$. The geometric features of two pedestrians are illustrated in Figure 2.1. In practical simulation, an anisotropic formula of the social-force is widely applied where Equation (2.3) is scaled by a function of $\lambda_i$. The angle $\varphi_{ij}$ is the angle between the direction of the motion of pedestrian $i$ and the direction to pedestrian $j$, which is exerting the repulsive force on pedestrian $i$. If $\lambda_i = 1$, the social force is symmetric and $0 < \lambda_i < 1$ implies that the force is larger in front of a pedestrian than behind. This anisotropic formula assumes that pedestrians move forward, not backward, and thus we can differ the front side from the backside of pedestrians based on their movement. Although the anisotropic formula is widely used in pedestrian modeling, it also brings a controversial issue that the anisotropic formula of social force disobeys Newton's 3rd law.

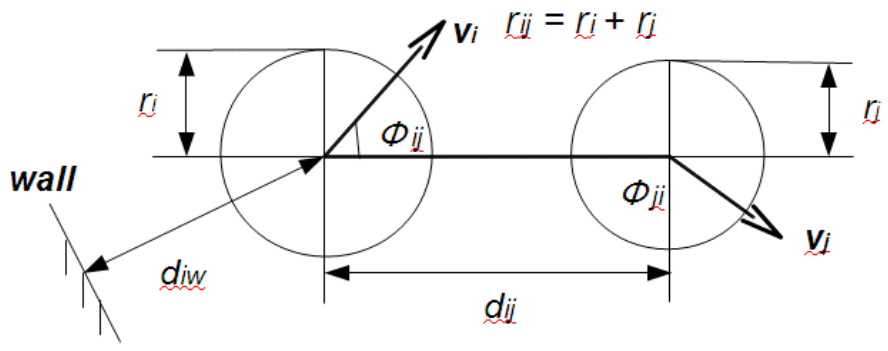

**Figure 2.1 A Schematic View of Two Pedestrians (Equation 2.3): The distance of pedestrians *i* and *j* is denoted by $d_{ij}$ and the sum of their radii is given by $r_{ij}$. Their interaction force is given by Equation 2.3.**

The physical interaction $\boldsymbol{f}_{ij}^{phy}$ describes the physical interaction when pedestrians have body contact, and it is composed by an elastic force that counteracts body compression and a sliding friction force that impedes relative tangential motion of two pedestrians. Both of them are valid only when $r_{ij}>d_{ij}$. In Helbing, Farkas and Vicsek, 2000 the interaction force is repulsive. The model may also include an attraction force in its original version (Helbing and Molnar, 1995, Helbing et al., 2002 Korhonen, 2017). The interaction of a pedestrian with obstacles like walls is denoted by $\boldsymbol{f}_{iw}$ and is treated analogously, i.e., $\boldsymbol{f}_{iw} = \boldsymbol{f}_{iw}^{soc} + \boldsymbol{f}_{iw}^{phy}$. Here $\boldsymbol{f}_{iw}^{soc}$ is also an exponential term and $\boldsymbol{f}_{ij}^{phy}$ is the physical interaction when pedestrians touch the wall physically.

So far all the forces as introduced above are deterministic. In the model there is also a small fluctuation force $\boldsymbol{\xi}_i$, which causes randomness in motion. This is fundamentally different from $\boldsymbol{f}_i^{drv}$ with deterministic direction and magnitude, and the fluctuation force $\xi_i$ is random in nature, and it is mainly inspired by stochastic thermal movement in molecular dynamics. The direction and magnitude of $\xi_i$ follow certain probability distributions such as normal distribution or uniform distribution. The correlation of $\xi_i$ at different time points is given by $cor(\xi_i(t) \cdot \xi_i(t'))=2S\delta(t-t')$, which means it is only self-correlated along the time axis. Based on statistical physics $\xi_i(t)$ is a temperature-related, and thus increasing the average fluctuation implies a kind of heating effect on the crowd.

By simulating many such individuals in collective motion, several scenarios in crowd movement were demonstrated, and one is called the “faster-is-slower” effect. This scenario was observed when a crowd pass a bottleneck doorway, and it shows that increase of desired speed (i.e., $|\boldsymbol{v}_i^0(t)|$) can inversely decrease the collective speed of crowd passing through the doorway. Another paradoxical phenomenon is called “freezing-by-heating,” and it studies two groups of people moving oppositely in a narrow passageway, and the simulation shows that increasing the fluctuation force in Equation (2.1) can also cause blocking in the counter-flow of pedestrian crowd. Other spatio-temporal patterns include herding effect, oscillation of passing directions, lane formation, dynamics at intersections and so forth.

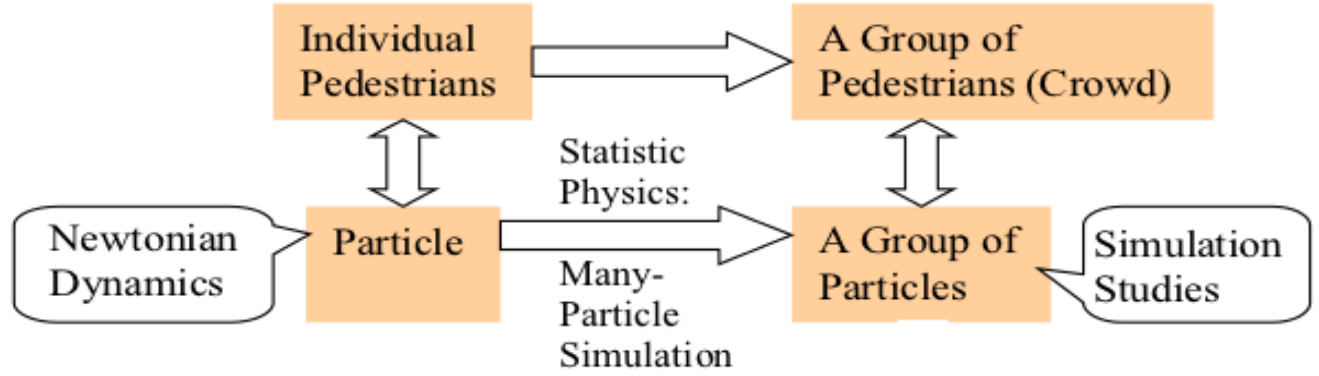

**Figure 2.2 Framework in Helbing, Farkas, and Vicsek, 2000: The method is generally from statistic physics, where each agent is a Newtonian particle and they interact with forces and emerge certain phenomena at the macroscopic level. This approach is consistent with agent-based simulation in field of computer science or complex systems.**

In the past decade, the social-force model has generated considerable research on evacuation modeling (Helbing and Johansson, 2010), and it has been incorporated into several egress simulators, such as Fire Dynamics Simulator with Evacuation (Korhonen et al., 2008, Korhonen, 2017), Pedsim, SimWalk, MassMotion (Oasys, 2018), VisWalk and Maces (Pelechano and Badler, 2006) and Menge. The model has been partly validated based on data sets from real-world experiments. The method of validation involves comparing the simulation of the model with associated observations drawn from video-based analysis (Johansson, Helbing and Shukla, 2007; Johansson et al., 2008).

A criticism about the model, as mentioned before, is that the anisotropic formula of social force disobeys Newton's 3rd law. If the model does not obey Newton 3rd law, it becomes questionable in field of physics studies. Another problem about the social force is the dilemma of choosing proper parameters to avoid both overlapping and oscillations of the moving pedestrians (Chraibi, et al., 2011). Certain scenarios about oscillation of walking behavior are not realistic in the physical world. From the perspective of physics these problems are difficult to explain. However, as mentioned previously, human motion is self-driven and self-adapted, and it is subject to both physical laws and psychological principles. Psychological study will give a new perspective to understand the model and provide us with a new angle to understand the problems. This is why we want to bridge the gap between psychological studies and physical explanation about this model.

## III. PSYCHOLOGICAL EXPLANATION OF SELF-DRIVING FORCE AND SOCIAL-FORCE

This section provides a psychological perspective to understand self-driving force and social-force, and the conception of stress is critically involved. Very importantly we introduce the new concept of desired distance in the social force, which is the counterpart of the desired velocity in the self-driving force, and this innovation is the foundation of our discussion in the next few sections.

### 3.1 Panic, Stress and Time-Pressure

One problem about the social-force model is that most of the testing results were explained by “panic” behavior of people (Helbing, Farkas and Vicsek, 2000; Helbing et al., 2002; Helbing and Johansson, 2010) while existing egress research clarifies the psychological state of panic occurs relatively rarely in real-world evacuation events (Sime, 1980; Proulx, 1993; Ozel, 2001), and this could cause misunderstanding of the model by social psychologists. Defined psychologically, “panic” means a sudden over-whelming terror which prevents reasoning and logical thinking, and thus results in irrational behavior. Based on Equation (2.1) and (2.2), we see that the equations do not imply any irrational behavior aroused by fear, but describe a kind of rational mechanics that govern an individual's motion. Thus, we think that the general use of the term panic is not essential to the social-force model.

By searching in literature of social psychological studies in emergency egress, we think that "stress" is more accurate conceptualizations of the social-force model than "panic." (Sime, 1980; Ozel, 2001). Psychological stress can be understood as the interaction between the environment and the individual (Selye, 1978, Staal, 2004), emphasizing the role of the individual's appraisal of situations in shaping their responses. In Stokes and Kite, 2001, such stress is the result of mismatch between psychological demand and realistic situation, and Equation (2.2) characterizes the mismatch in terms of velocity: the psychological demand is represented by desired velocity $\boldsymbol{v}_i^0$ while the physical reality is described by the physical velocity $\boldsymbol{v}$. The gap of two variables describes how much stress people are bearing in mind, and thus are motivated into certain behavior in order to make a change in reality. Such behavior is formulated as the self-driving force in Equation (2.1) and (2.2).

Furthermore, velocity is a time-related concept in physics and the gap of velocities actually describes a kind of time-related stress, or commonly known as time-pressure. Such a kind of stress is caused by insufficient time when people are dealing with a time-critical situation, and time is the critical resource to complete the task. In sum, although the social-force model is labeled with the term "panic," its mathematical description is not directly related to "panic" in a psychological sense and the self-driving force critically characterizes the psychological concept of stress and time-pressure. This also explains why the model can be well used in simulation of emergency egress because "emergency" implies shortage of time in a process.

### 3.2 Interpersonal Stress and Social-Force

As above we briefly discuss the time-related stress and explain its relationship to desired velocity. There is another kind of stress originating from social relationship, and it leads to competition or cooperation in crowd movement, and such stress is characterized by the social-force. This subsection will discuss such stress and its relationship with interpersonal distance, and we will introduce a new concept of desired distance in the social force, and it is the counterpart of desired velocity in the self-driving force.

Interpersonal distance refers to a theory of how people use their personal space to interact with surrounding people. In Hall, 1963 the theory was named by proxemics, and it was defined as "the interrelated observations and theories of man's use of space as a specialized elaboration of culture." Proxemics suggests that we surround ourselves with a "bubble" of personal space, which protects us from too much arousal and helps us keep comfortable when we interact with others. Psychologically, proximal distance origins from basic human instincts that we define a personal space to get a sense of safety. People normally feel stressed when their personal space is invaded by others. There are four interpersonal distances mentioned in Hall, 1966: intimate distance (<0.46m), personal distance (0.46m to 1.2m), social distance (1.2m to 3.7m), and public distance (>3.7m), and each one represents a kind of social relationship between individuals. Here we highlight two issues about proxemics as below.

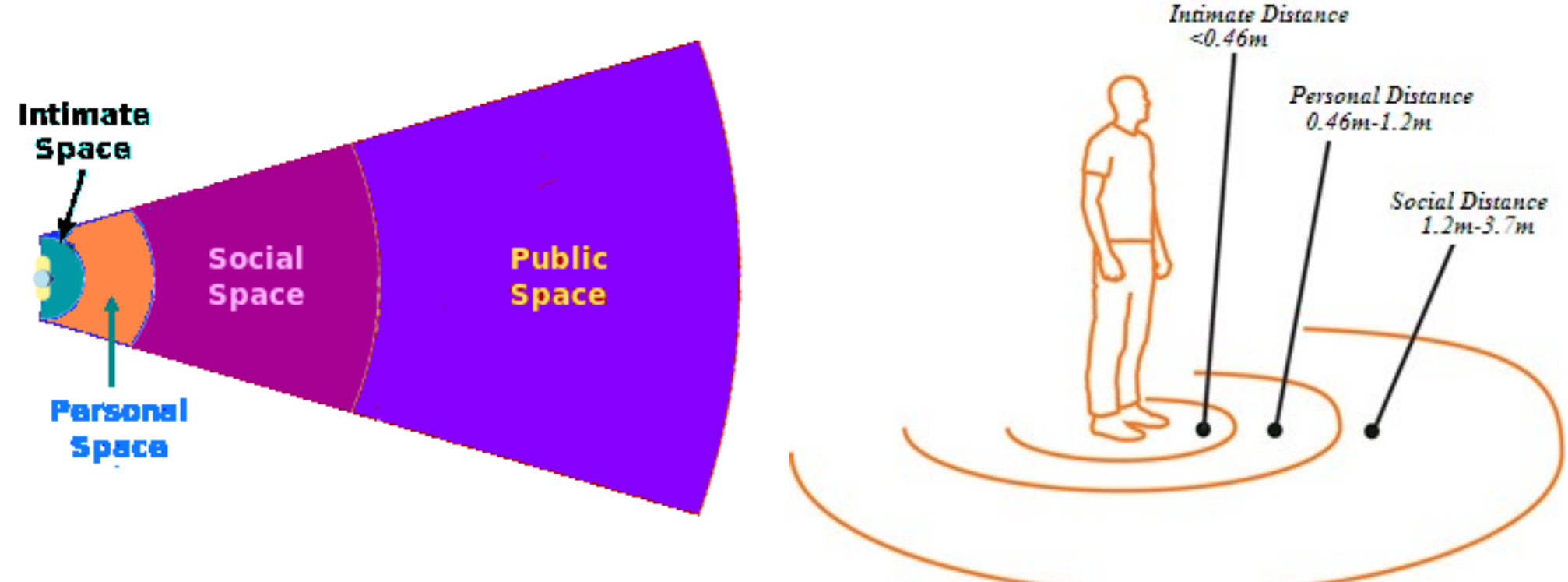

**Figure 3. Description of Proxemics as Introduced in Hall, 1963: Proximal distance origins from basic human instincts that we define a personal space to get a sense of safety. People normally feel stressed when their personal space is invaded by others. There are four interpersonal distances mentioned in Hall, 1966: intimate distance (<0.46m), personal distance (0.46m to 1.2m), social distance (1.2m to 3.7m), and public distance (>3.7m).**

(a) The interpersonal distance is object-oriented. For example, we usually keep smaller distance to a friend than to a stranger, and such distance is an indication of familiarity. As named by personal distance (0.46m to 1.2m) in proxemics, this range is widely observed as the distance to interact with our friends or family, and normal conversations take place easily at this range. In contrast when moving with strangers in a crowd, for example, in a shopping mall or railway stations, we usually keep social distance (1.2m to 3.7m) with surrounding people.

(b) The interpersonal distance is occasion-dependent, which reflects a kind of social norms and cultures. For instance, in a crowded train or elevators, although such physical proximity is psychologically disturbing and uncomfortable, it is accepted as a social norm of modern life. Moreover, proxemics is also culture-dependent, and a well-known example is that Muslim culture defines larger distance between the male and female on public occasions than modern west cultures. In brief although proximal distance origins from basic human instincts, it is also redefined in different social norms and cultures.

Proxemics implies that when the interpersonal distance is smaller than the desired, people feel stressed. Repulsion comes into being in this situation, and repulsion increases when the distance further decreases. This theory justifies the assumption of repulsive social-force in Equation (2.3). However, the repulsion is not related to physical size of two people (i.e., $r_{ij}$), but the social relationship, culture and occasions. Comparing social force with self-driving force, we suggest that there should be a subjective concept of desired distance $d_{ij}^0$ in the social force, and it replaces $r_{ij}$ in Equation (2.3). Here $d_{ij}^0$ is the target distance that individual $i$ expects to keep with individual $j$, and it describes social relationship of individual $i$ and $j$. By using the exponential form in Equation (2.3), the social force is rewritten as

$$\boldsymbol{f}_{ij}^{soc} = A_i \exp\left[\frac{(d_{ij}^0 - d_{ij})}{B_i}\right]\boldsymbol{n}_{ij} \quad \text{or} \quad \boldsymbol{f}_{ij}^{soc} = \left(\lambda_i + (1-\lambda_i)\frac{1+\cos\varphi_{ij}}{2}\right) A_i \exp\left[\frac{(d_{ij}^0 - d_{ij})}{B_i}\right]\boldsymbol{n}_{ij} \tag{3.1}$$

**Table 3.1 On Conception of Stress in Social-Force Model**

| | *Opinion (Psychological Characteristics)* | *Behavior (Physic-Based Characteristics)* | *Difference between subjective opinion and objective reality* |
|---|---|---|---|
| ***Time-Related Concept: Velocities*** | desired velocities $\boldsymbol{v}_i^0 = v_i^0 \boldsymbol{e}_i^0$ | actual velocities $\boldsymbol{v}_i = v_i \boldsymbol{e}_i$ | ***Time-Related Stress:*** $\boldsymbol{v}_i^0 - \boldsymbol{v}_i$ |
| ***Space-related Concept: Distances*** | desired distance $d_{ij}^0$ | actual distance $d_{ij}$ | ***Space-related Stress:*** $d_{ij}^0 - d_{ij}$, |

Similar to desired velocity $\boldsymbol{v}_i^0$, the desired distance $d_{ij}^0$ is the target distance in one's mind, specifying the distance that one expects to adapt oneself with others. The physical distance $d_{ij}$ is the distance achieved in the reality. The gap of $d_{ij}^0$ and $d_{ij}$ implies the difference between the subjective wish in one's mind and objective feature in the reality. Similar to $\boldsymbol{v}_i^0 - \boldsymbol{v}_i$, as an indication of time-related stress concerning emergencies, $d_{ij}^0 - d_{ij}$ is an indication of interpersonal stress related to the social space. Such stress depends on the intrinsic social characteristics of the crowd, not directly related to the emergency situation. Here $A_i$ and $B_i$ are parameters as introduced before, and $\boldsymbol{n}_{ij}$ is the normalized vector which points from pedestrian $j$ to $i$. In a similar manner, an anisotropic formula of the social-force is also modified in Equation (3.1). The social force also functions in a feedback manner to make the realistic distance $d_{ij}$ approaching towards the desired distance $d_{ij}^0$. A difference is that $\boldsymbol{v}_i^0$ and $\boldsymbol{v}_i$ are vectors while $d_{ij}^0$ and $d_{ij}$ are scalars.

Although a major difference exists between the concepts of $r_{ij}$ and $d_{ij}^0$, most of the simulation results in Helbing, Farkas and Vicsek, 2000 still stand. In fact, the coding framework of social-force model is not affected when $r_{ij}$ is replace by $d_{ij}^0$. When realizing the model in computer programs, $r_{ij}$ and $d_{ij}^0$ are exactly at the same position in coding work, and we can simply have $d_{ij}^0 = r_{ij} \cdot c_{ij}$ to extend the traditional social force to the new force, and $c_{ij}>1$ is a scale factor.

As above we modify the social force by introducing desired interpersonal distance $d_{ij}^0$. In a similar way the wall repulsion could be modified as well by introducing desired wall distance $d_{iw}^0$, and $d_{iw}^0$ is the target distance that individual $i$ expects to keep with surrounding walls as shown in Table 3.2. For more mathematical details, please refer to Wang, 2020.

In a psychological sense $d_{ij}^0$ and $\boldsymbol{v}_i^0$ are both subjective concepts which exist in people's mind, and they characterize how an individual intends to interact with surroundings. As a result, the social-force given by Equation (3.1) and the driving force are both subjective entities, which are self-controlled by one's intentions and opinions. The anisotropic formula of social force also takes human foresight into account such that each individual is more influenced by things in front than behind. Thus, the anisotropic formula assumes that human has perception to the surrounding things, and it is also a subjective process involving human cognition and mental activities. In this paper we will differ such self-controlled forces from common forces in classic physics. The common forces are passively received from surroundings while the social force and driving force are not, but intentionally generated by an individual. In brief the social force is not a physics thing, but a subjective concept, which describes one's opinions and mental activities on how an individual desires to interact with surrounding people.

In a physics sense the social force is generated by the foot-ground friction, which exactly obey physics laws. In fact two individuals commonly do not have any physical interaction if they do not touch each other physically. If people want to keep a proper distance with others, they do not need to implement any forces on others, but simply move on foot to adjust the distance. In other words, we normally use foot-ground friction to realize such "social force." Here the foot-ground friction is a physics concept, and it agrees with Newton 3rd Law for interaction between foot and ground. We cannot move in a world without friction, and Newton 3rd Law governs our motion definitely. In a general sense when subjective forces in our mind are realized into the physical world, they must become certain physics-based forces and follow physics laws. For example, when the social force and self-driving force are realized in pedestrian motion, they become part of foot-ground friction, and it follows Newton Laws, and pedestrians can consciously control this friction to decide where and how fast to move. Thus such subjective forces shows a connection between the psychological principles of human mind and physical world of universe.

**Table 3.2 About Newton's Laws and Social Force Model**

<table>
<tr><td colspan="3">$m_i \frac{d\boldsymbol{v}_i(t)}{dt} = \boldsymbol{f}_i^{drv} + \sum_{j(\neq i)} (\boldsymbol{f}_{ij}^{soc} + \boldsymbol{f}_{ij}^{phy}) + \sum_{w} (\boldsymbol{f}_{iw}^{soc} + \boldsymbol{f}_{iw}^{phy}) + \boldsymbol{\xi}_i$</td></tr>
<tr><td>Self-Driving Force</td><td>$\boldsymbol{f}_i^{drv} = m_i \frac{\boldsymbol{v}_i^{\boldsymbol{0}}(t) - \boldsymbol{v}_i(t)}{\tau_i}$</td><td rowspan="3">These forces are generated by intentions of people and are realized by the foot-ground friction, which exactly obey Newton's laws.</td></tr>
<tr><td>Social Force</td><td>$\boldsymbol{f}_{ij}^{soc} = A_i \exp\left[\frac{(d_{ij}^0 - d_{ij})}{B_i}\right] \boldsymbol{n}_{ij}$</td></tr>
<tr><td>Wall Repulsion (Psychological Component)</td><td>$\boldsymbol{f}_{iw}^{soc} = A_{iw} \exp\left[\frac{(d_{iw}^0 - d_{iw})}{B_{iw}}\right] \boldsymbol{n}_{iw}$</td></tr>
</table>

In social force model the foot-ground friction mainly consists of self-driving force and social force as shown in Table 3.2, and we assume that vectorial additivity of these force components which represents different environmental influence on the moving pedestrian. This assumption is acceptable because human mind has the capability of parallel perceiving and thinking, which means people are able to adjust their interpersonal distance (i.e., social force) while still keep their destination in mind (i.e., self-driving force). If an individual is very close to another, the social force will increase such that it becomes predominant in the joint force. In contrast if an individual is not surrounded with other individuals or facilities, the self-driving force predominate the joint force such that one will just head to the destination. Thus, the vectorial additivity of subjective forces is a reasonable assumption for pedestrian modeling at current stage, and it enables us to use vectorial math to model opinions of pedestrians.

### 3.3 Faster-is-Slower Effect and Yerkes–Dodson law

In this subsection we would like to further justify the social-force model by the concept of stress, and we will investigate a typical scenario of crowd evacuation. This scenario was named by "faster-is-slower" effect in Helbing, Farkas and Vicsek, 2000, and it refers to egress performance when a large number of individuals pass through a narrow doorway. The simulation result shows that the egress time may inversely increase if the average desired velocity keeps increasing. In other words, egress performance may degenerate if the crowd desire moving too fast to escape. We will explain the simulation result from the psychological perspective of stress and time-pressure. In particular, this scenario reiterates an existing psychological knowledge: moderate stress improves human performance (i.e., speeding up crowd motion); while excessive stress impairs their performance (i.e., disorders and jamming), and this theorem is widely known as Yerkes–Dodson law in psychological study (Yerkes and Dodson, 1908; Teigen, 1994; Wikipedia, 2016). In addition, this subsection will keep using psychological theory to understand the simulation result of the social force model, and if readers are not quite interested in the doorway scenario, you can omit this section and move on to the next section without any problems.

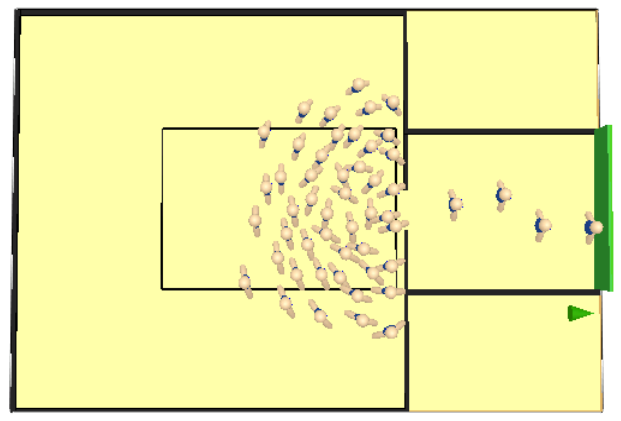

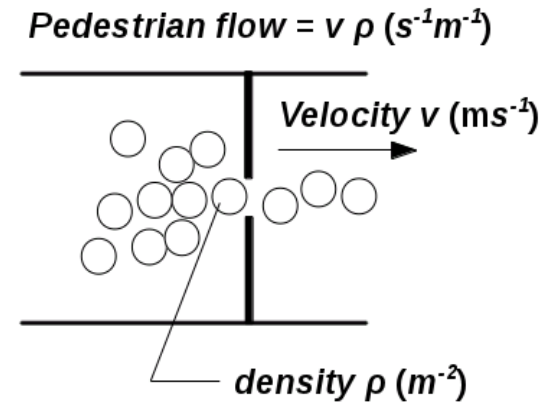

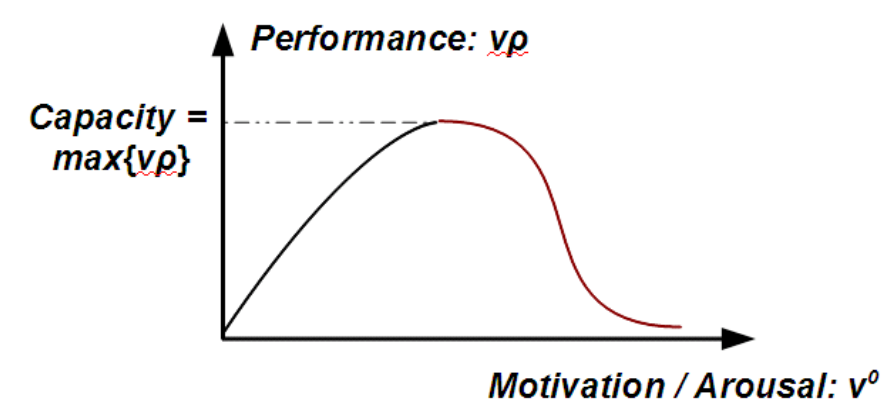

**Figure 3.1 About crowd movement in a passageway: The simulation result in Helbing, Farkas and Vicsek, 2000 shows that the egress time may inversely increase if the average desired velocity keeps increasing. In other words, egress performance may degenerate if the crowd desire moving too fast to escape. This finding is consistent with an existing psychological principle, Yerkes–Dodson law, which states the relationship between arousal level and performance: performance increases with arousal, but only up to a point. Beyond the point the arousal becomes excessive and the situation is much stressful such that performance diminishes.**

Yerkes–Dodson law states the relationship between arousal level and performance: performance increases with arousal, but only up to a point. Beyond the point the arousal becomes excessive and the situation is much stressful such that performance diminishes. The arousal level indicates the intensity of motivation and it depends on stimulus strength from environment (e.g. alarm or hazard). Motivation leads to behavioral response. In the social-force model, the arousal or motivation is represented by desired velocity $\boldsymbol{v}^0$, and the behavioral response is represented by actual velocity $\boldsymbol{v}$. The performance of crowd escape is measured by pedestrian flow $\rho v$ at the doorway, describing how many individuals pass through a doorway of unit width per time unit (See Figure 3.1, $\rho$ and $v$ are the crowd density and physical speed nearby the doorway). The pedestrian flow is limited by the passage capacity, which determines the maximal pedestrian flow that people are able to realize in collective motion (Wang et al., 2008). In other words, the passageway capacity determines the critical point in Yerkes–Dodson law, indicating whether the collective motivation is excessive or not.

(a) When the passage capacity is sufficient, $v$ increases along with $v^o$ while $\rho$ can be adjusted such that the physical distance among people is psychologically comfortable. As a result, people are able to move as fast as desired while still keep proper interpersonal distance. This scenario corresponds to the increasing segment of the curve in Figure 3.1, and we can call it faster-is-faster effect in this paper. This effect has been partly observed in real-world experiments in Daamen and Hoogendoorn, 2012[1], and it is also justified by the simulation result in Høiland-Jørgensen et al., 2010. Although the phenomenon was not emphasized in Helbing, Farkas, and Vicsek, 2000, we can still infer this effect from the testing result in the paper[2].

(b) When the passage is saturate, the physical speed $v$ and density $\rho$ reach the maximum and the pedestrian flow $\rho v$ is the maximal. In this situation further increasing $v^o$ will compress the crowd and increase the repulsion among people. As the repulsion increases, the risk of disorder and disaster at the bottleneck increases correspondingly (e.g., jamming and injury). If such disastrous events occur, the moving crowd will be significantly slowed down and the faster-is-slower effect comes into being, and this corresponds to the decreasing segment of the curve in Figure 3.1.

In sum, as motivation level $v^0$ increases, there are two scenarios as introduced above. The relationship between $v^0$ and performance $\rho v$ is depicted by an inverted-*U* curve as shown in Figure 3.1. Here the motivation level $v^0$ especially depends on environmental stressors, which are any event or stimulus perceived as threats or challenges to individuals. For example, in emergency evacuation a sort of important stressors are hazardous condition (e.g., fire and smoke). Perception of hazard will increase arousal level so that desired velocity $v^0$ increases. In addition, whether faster-is-slower effect emerges also depends on whether people tend to compete or cooperate with each other. To be consistent with real-world observation we think a falling-down model is needed to exhibit the faster-is-slower effect, and people fall down mainly due to physical contact (e.g., pushing) when competitive behavior is intensified. If someone falls down at the bottleneck, faster-is-slower phenomenon will emerge and pedestrian flow $\rho v$ will decrease sharply (Wang, 2020). In other words, the competitive behavior and falling-down model are important issues, and the desired interpersonal distance $d_{ij}^0$ also plays an important role.

---

[1] In Daamen and Hoogendoorn, 2012 the crowd are excited and the designed experiment does not typically refer to the stress and time-pressure. Exciting the crowd is similar to heating the particles, which can also cause jamming in the bottleneck ("freezing-by-heating effect in Helbing et al., 2002). However, this effect can be considered as an extension of the faster-is-slower effect. Please refer to Section 6.2 for more details.

[2] In Helbing, Farkas, and Vicsek, 2000 the simulation results demonstrated how the desired velocity $v^d$ affects the "physical flow $\rho v$ divided by the desired velocity" in order to emphasize the faster-is-slower effect (See Figure 1(d) in Helbing, Farkas, and Vicsek, 2000). Also, there is an ascending portion of the curve in Figure 1(d) as the desired velocity increases initially, and this portion actually exhibited the faster-is-fast effect. We infer that the ascending of the curve will be more significant if the physical flow $\rho v$ ($s^{-1}m^{-1}$) is directly plotted.

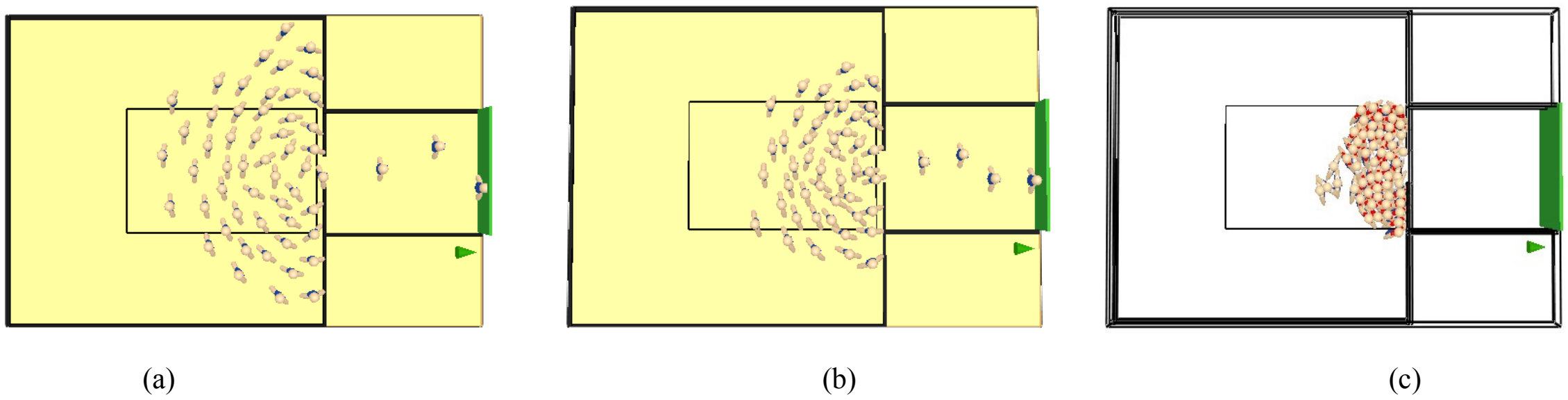

(a) (b) (c)

**Figure 3.2 About Social Force and Faster-Is-Slower Effect: (a) Use large $d_{ij}^0$ in normal situation such that people obey social norm of large interpersonal distance. The result is decrease of flow rate and less chance of physical interaction. (b) Use small $d_{ij}^0$ in emergency egress such that people follow the social norm of small interpersonal distance. Flow rate thus increases and physical interaction increase in a stochastic sense. (c) As $d_{ij}^0$ continues to decrease, the physical interaction causes someone to fall down, and the doorway is thus blocked by those falling-down people.**

To investigate how $d_{ij}^0$ affects the pedestrian flow at bottlenecks, we implement the following test by slightly modifying the source program of FDS+Evac. The test is based on IMO door flow test (IMO, 2007), where the door width is $1\,m$, and it is also the door width used in Helbing, Farkas and Vicsek, 2000. The left diagram corresponds to large $d_{ij}^0$, where we specify $d_{ij}^0 = 3 \cdot r_{ij}$ , while the middle diagram corresponds to relatively small $d_{ij}^0$, where $d_{ij}^0 = 2 \cdot r_{ij}$ is used. Here $r_{ij}$ is the sum of the radii of individual $i$ and $j$, namely, $r_{ij} = r_i + r_j$ (See Figure 2.1). The comparative results suggest that decreasing desired distance $d_{ij}^0$ moderately will increase pedestrian flow rate at the doorway. If desired distance $d_{ij}^0$ keeps decreasing as shown in the right diagram, physical interaction may become predominant. If such physical contact is a kind of competitive behavior (e.g, pushing or shoving), someone may fall down due to such physical contact. As a result the doorway will be blocked, and the flow rate will reduce sharply. Thus, we think that the relationship between $d_{ij}^0$ and performance $fv$ is also depicted by an inverted-U curve, which exactly follow the Yerkes–Dodson law also.

In sum, mismatch of psychological demand and physical reality results in a stressful condition. In emergency egress, such stress is aroused from environmental factors in two categories. A major kind of factors include hazard conditions and alarm, resulting in impatience of evacuees and causing time-pressure. The psychological model refers to the fight-or-flight response (Cannon, 1932), where hazardous stimulus motivate organisms to flee such that the desired velocity increases. Another kind of stress is aroused from surrounding people, resulting in interpersonal stress in collective behavior. Such stressor makes one repulsive with others, and it determines whether people tend to compete or cooperate with each other. In brief, the simulation of the faster-is-slower effect reiterate Yerkes–Dodson law with respect to two-dimensional stressors.

From the perspective of stress Yerkes–Dodson law is also understood by dividing stress into eustress and distress (Selye, 1975): stress that enhances function is considered eustress. Excessive stress that is not resolved through coping or adaptation, deemed distress, may lead to anxiety or withdrawal behavior and degenerate the performance. Thus, stress could either improve or impair human performance. Traditionally, this psychological theorem mainly refers to performance at individual level, such as class performance of a student or fight-or-flight response of an organism. The simulation of social-force model reiterates this well-known psychological knowledge in collective behavior. In brief, the testing result of social-force model agrees with Yerkes–Dodson law and it provides a new perspective to understand this classic psychological principle.

## IV. HERDING EFFECT AND GROUP DYNAMICS

Stress is perceived when we think the demand being placed on us exceed our ability to cope with, and it can be external and related to the environment, and it becomes effective by internal perceptions. The motivation level $v_i^0$ and $d_{ij}^0$ are the result of such perception, and are adapted to the environmental stressors. In psychology theory of behaviorism this process is called stimuli-reaction process (S-R process). As a result, stress refers to human response and adaption to the environment, and it is feasible to extend social-force model to characterize the interplay between individuals and their surroundings. As below we present a diagram to describe the interplay between individuals and their surroundings based on the extended social-force model.

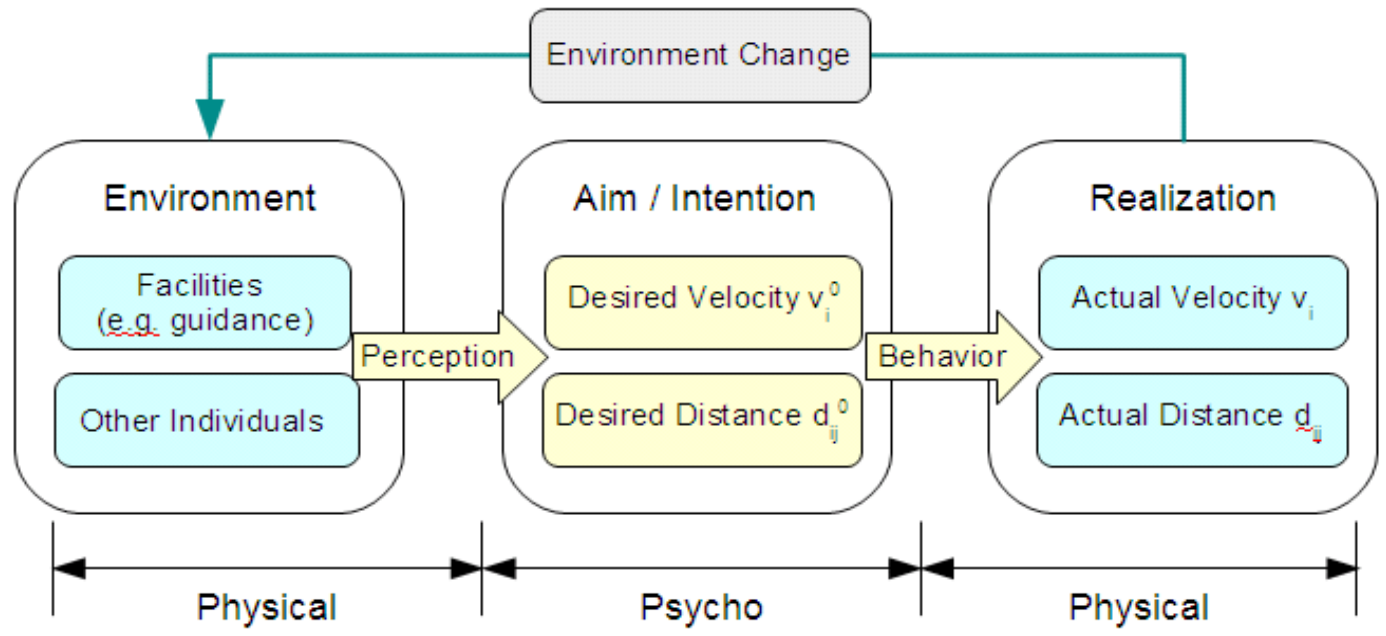

**Figure 4.1 Perception and Behavior in a Feedback Mechanism: The pedestrian motion is a response to stressors in environmental conditions ((e.g, hazard, exit signs)), and $\boldsymbol{v}_i^0$ and $d_{ij}^0$ could vary both temporally and spatially, and lead to change of physical variables $\boldsymbol{v}_i$ and $d_{ij}$.**

In the above diagram environmental factors include surrounding people and facilities (e.g., alarm, guidance and hazard in egress). The resulting pedestrian motion is a response to stressors in environmental conditions, and $\boldsymbol{v}_i^0$ and $d_{ij}^0$ could vary both temporally and spatially. As for the facilitates (e.g, hazard, exit signs), we will briefly explain how to apply the above model in simulation of crowd evacuation as below.

(a) About Hazard: When hazard propagates towards people, people normally desire moving faster to escape from danger or search for familiar ones to escape together. As a result, the desired velocity $\boldsymbol{v}^0$ increases in order to escape from danger while $d_{ij}^0$ will decrease in order to make people cohesive to stay together. The direction of $\boldsymbol{v}^0$ is pointing to the place of safety where the hazard level decreases, but it also depends on the high-level decision making process such as exit choice in egress.

(b) Guidance: Guidance such as exit signs will guide people to place of safety and it changes the destination of escape. The guidance thus affects the direction of desired velocity $\boldsymbol{v}_i^0$ with certain probability. Existing research shows that people tend to use the familiar path and trust more on personalized guidance (Kuligowski, 2009). Such factors are integrated into a probabilistic graph model (Wang et al., 2008) to affect the direction of desired velocity.

(c) Social Norms: In the field of social psychology, social norms are defined as "representations of appropriate behavior" in a certain situation or environment. From the perspective of crowd modeling, the social norm is indicated by $d_{ij}^0$. For example in elevators or entrance of a passageway, people commonly accept smaller proximal distance. Thus, the desired interpersonal distance is smaller there, and $d_{ij}^0$ is to be scaled down proportionally in these places. In brief, $d_{ij}^0$ is occasion-dependent, and it varies along with locations. In emergencies the social norm is modified such that competitive behavior may emerge, and the model is thus useful to investigate crowd behavior in emergencies.

In addition to surrounding entities such as hazard, alarm and guidance (e.g., exit signs), another important factor is human factor. That is, how an individual's opinion and behavior are affected by surrounding people, and this issue leads to grouping dynamics and self-organizing phenomenon in pedestrian crowd and we will next elaborate this issue in detail.

## 4.1. Herding Behavior and Opinion Dynamics

Herding is especially evident when people are responding to an emergency (Low, 2000). Emergency implies time-pressure as mentioned before and excessive time-pressure weakens the ability of logical thinking and reasoning, and independent decision making is more difficult in stressful conditions. Thus, people are more inclined to follow others (e.g., neighbors' decisions) rather than make decisions by themselves. Based on the social-force model in Helbing, Farkas and Vicsek, 2000, the computational method of herding behavior is presented with certain extension as below.

$$\boldsymbol{e}_i^0(t+1) = Norm[(1-p_i)\boldsymbol{e}_i^0(t) + p_i[\boldsymbol{e}_{others}(t)]_{\to i}] \qquad [\boldsymbol{e}_{others}(t)]_{\to i} = \Sigma_{R_i}\boldsymbol{e}_j(t) \tag{4.1}$$

$$v_i^0(t+1) = (1-p_i)v_i^0(t) + p_i[v_{others}(t)]_{\to i} \qquad [v_{others}(t)]_{\to i} = (\Sigma_{R_i} v_j(t))/N \tag{4.2}$$

where the magnitude of velocity is the moving speed, i.e., $\boldsymbol{v}_i^0 = v_i^0 \boldsymbol{e}^0_i$ and $\boldsymbol{v}_j = v_j \boldsymbol{e}_j$

The desired direction $\boldsymbol{e}^0_i$ is updated by mixing itself with the average direction of others within radius $R_i$. Norm[] represents normalization of a vector. Both options are weighted with some parameter ($1-p_i$) and $p_i$, and two options follow two-point distribution with probability ($1-p_i$) and $p_i$, and $e^0_i$ is updated by the statistical average. As a consequence, individualistic behavior is dominant if $p_i$ is low whereas herding behavior dominates if $p_i$ is high. The mixture of two choices also exists for desired moving speed in a similar manner. For example, if one's neighbors all move fast, he or she may also wants to speed up to carry on with others. Thus, Equation (4.2) is presented to describe how an individual's desired speed interacts with others in collective behavior (Lakoba, Kaup and Finkelstein, 2005).

In sum the above equations describe how desired velocity $\boldsymbol{v}_i^0 = v_i^0 \boldsymbol{e}^0_i$ is updated, and the model refers to existing theories in opinion dynamics (Deffuant, 2000). Based on Equation (4.1) and (4.2) we know how an individual's velocity is affected by surrounding ones in terms of both direction and speed. Here we have the following remarks.

One may notice that $e^0_i$ is updated based on the sum of others' moving directions while $v^0_i$ is updated based on the arithmetic average of others' moving speed. The explanation is given as follows. Suppose there are $N$ people within radius $R_i$ moving towards a common destination, and the arithmetic average mean of $e_j(t)$ is the same as anyone among them. However, from the perspective of social-psychological viewpoint $N$=6 people should have more impact on one's opinion than $N$=1. The more source people are there, the more impact is on your opinion, and it is an established psychological finding which suggests that social impact within a social structure is affected by strength (S), immediacy (I) and number of people (N). This finding is named by the social impact theory developed by Bibb Latané in 1980's. As a result, Equation (4.1) and (4.2) suggests that the social impact increases with the number of source people acting on the target individual.

Thus, as for the desired moving direction we suggest that $e_j(t)$ should be updated by the sum rather than the average in order to reflect social impact effect. The effect is addible in a sense that the arithmetic average cannot measure such impact for it does not significantly increase with number of people $N$. In contrast, for the moving speed, it is not an addible effect. If an individual is surrounded by many others moving faster than him, the individual may also want to speed up his pace, and the desired speed refers to the average speed of others, not the sum. Thus, it is reasonable to update $v^0_i$ by the arithmetic average. This effect is well introduced in studying collective motion of bird flock (Vicsek et. al., 1995).

(a) Opinion and Behavior

Another question of Equation (4.1) is whether people should follow opinion or behavior of others. In psychology theory behavior and opinion are different but related concepts. Opinion is the cause of behavior, and thus we can infer one's opinion from his or her behavior. For example, if we observe an individual's behavior of moving toward an exit, we will infer his or her opinion of choosing to use the exit. Thus, it remains a question in Equation (4.1) whether to use desired directions $\boldsymbol{e}_j^0(t)$ or the actual moving directions $\boldsymbol{e}_j(t)$ to aggregate others' characteristics. A similar question also exists for moving speed, that is, whether to use $v_j^0(t)$ or $v_j(t)$ in Equation (4.2). In general people are able to learn from others via talking or observing their behavior, and their own opinions are thus influenced. If we assume that an individual observes what others are doing and follow them, we should use $\boldsymbol{e}_j(t)$ and $v_j(t)$ in Equation (4.1) and (4.2). If we assume opinion are exchanged by language, $\boldsymbol{e}_j(t)$ and $v_j(t)$ will be replaced by $\boldsymbol{e}_j^0(t)$ and $v_j^0(t)$.

Table 4.1. Opinion and Behavior

| | | |
|---|---|---|
| Opinion (Psychological Characteristics) | desired velocities $\boldsymbol{v}_i^0 = v_i^0 \boldsymbol{e}^0_i$ | desired distance $d_{ij}^0$ |
| Behavior (Physic-Based Characteristics) | actual velocities $\boldsymbol{v}_i = v_i \boldsymbol{e}_i$ | actual distance $d_{ij}$ |

From the perspective of system dynamics $\boldsymbol{v}_j(t)=v_i \boldsymbol{e}_i$ should converge to $\boldsymbol{v}_j^0(t)=v_i^0 \boldsymbol{e}^0_i$ with a time delay, meaning that behavior is guided from opinion. This dynamics is consistent with the psychological viewpoint of behaviorism, which claimed that there is no scientific foundation to directly study human mind, but we can indirectly explore human mind from their behavior. In other words human behavior implies their opinion, and there is inherent relationship between these two subjects.

In our model the desired velocity $\boldsymbol{v}_i^0$ and physical velocity $\boldsymbol{v}_i$ are an example of such inherently related entities, and $\boldsymbol{v}_j(t)$ is expected to converge to $\boldsymbol{v}_j^0(t)$ within a relaxation time period. If such convergence is realized smoothly as in most normal cases, the relaxation time period is relatively short, it does not matter whether using $\boldsymbol{v}_j(t)$ or $\boldsymbol{v}_j^0(t)$ to to aggregate $[\boldsymbol{v}_{others}(t)]_{\rightarrow i}$ or $[\boldsymbol{e}_{others}(t)]_{\rightarrow i}$ in Equation (4.1) and (4.2) because the $\boldsymbol{v}_j(t)$ converges to $\boldsymbol{v}_j^0(t)$ quickly. Recall Equation (2.1) and (2.2) the time parameter ' is in charge of tuning the relaxation time, and the relaxation time reduces with decreasing ' in the model.

However, there do exist abnormal cases such that ' fails to tune the relaxation time. As a result, long relaxation time does exists, for example, as crowd compete to pass a bottleneck as shown in Figure 3.1, where $\boldsymbol{v}_j(t)=v_i \boldsymbol{e}_i$ cannot converge to $\boldsymbol{v}_j^0(t)=v_i^0 \boldsymbol{e}^0_i$ as fast as desired, and there is noticeable difference between using $\boldsymbol{v}_j(t)=v_i \boldsymbol{e}_i$ and $\boldsymbol{v}_j^0(t)=v_i^0 \boldsymbol{e}^0_i$ . In this case, if we use physical velocity $\boldsymbol{v}_j(t)=v_i \boldsymbol{e}_i$ to aggregate $[\boldsymbol{v}_{others}(t)]_{\rightarrow i}$ or $[\boldsymbol{e}_{others}(t)]_{\rightarrow i}$ , it means that individuals follow others to stand still, and thus $\boldsymbol{v}_i^0 = v_i^0 \boldsymbol{e}^0_i$ is expected to converge to zero. This setting implies that the whole crowd become less and less motivational, and every individual just desire to stand still by using Equation (4.2). This phenomenon-based setup contradicts with the underlying truth that people actually should feel more and more impatient in such a situation, and thus $\boldsymbol{v}_i^0$ is not expected to be reduced to zero, but should be kept constant or even increasing in the contrary. Therefore, we emphasize that it is not always reasonable to use behavior such as $\boldsymbol{v}_j$(t) and $\boldsymbol{e}_j$(t) to aggregate $[\boldsymbol{v}_{\text{others}}(t)]_{\to i}$ or $[\boldsymbol{e}_{\text{others}}(t)]_{\to i}$ . An important assumption is that opinions interact among individuals, and behavior is merely a media by which individuals acquire others' opinions. In other words, sometimes we cannot directly know other people's opinions or thoughts, and thus infer them directly from others' behavior. Thus, it is opinions that interacts ultimately, and this is a reasonable assumption in our dynamical model as well as in existing psychology theory. As below we introduce a generalized framework to describe exchange of opinions. The model is basically an extension of the well-known opinion dynamic process.

$$
\begin{aligned}
& opinion_i(t+1)=(1-p_i)\,opinion_i(t)+p_i[others(t)]_{\to i} \\
& [others(t)]_{\to i}=\Sigma_j\, c_{j\to i}\, opinion_j(t) \quad \Sigma_j\, c_{j\to i}=1 \\
\text{or}\ & [others(t)]_{\to i}=\Sigma_j\, c_{j\to i}\, c^{op}_{j\to i}\, opinion_j(t)+c_{j\to i}\, c^{be}_{j\to i}\, behavior_j(t) \quad \Sigma_j\, c_{j\to i}=1 \quad c^{op}_{j\to i}+c^{be}_{j\to i}=1
\end{aligned}
\qquad (4.3)
$$

where $c_{j\to i}$ is non-zero if individual $i$ is able to perceive or acquire opinion of individual $j$ such that individual $i$ is influenced by individual $j$. An issue to be emphasized is on the diagonal element of $C=[c_{j\to i}]_{n\times n}$ . We define $c_{i\to i}=0$ if individual $i$ is influenced by any other individuals in the social topology. If not, let the diagonal element $c_{i\to i}=1$ to meet the requirement of $\sum_j c_{j\to i}=1$. Thus, for practical computing, Equation (4.3) is applicable to either an isolated individual or an individual who is socially connected to others. As for an isolated individual who is not influenced by any others, it gives that $c_{j\to i}=0$ for $\forall j\neq i$, and we define $c_{i\to i}=1$ such that opinion$_i(t)$ keeps unchanged. This definition looks a little intricate, but it enables us to discuss how $C=[c_{j\to i}]_{n\times n}$ and $\Pi = diag(p_1, p_2, \ldots p_n)$ are dynamically changing as two separate issues.

Now it is feasible to further integrate parameter $p_i$ with matrix $C=[c_{j\to i}]_{n\times n}$ , describing how the opinion of individual $i$ is socially formed from matrix $P=[p_{ij}]_{n\times n}$. Based on Equation (4.3), matrix $P$ is given by $P = (I-\Pi) + \Pi\cdot C,$ where $\Pi = diag(p_1, p_2, \ldots p_n)$ is a n×n square matrix with its diagonal element defined by $p_1, p_2, \ldots p_n$ , and $I$ is a n×n the identity matrix. In other words, we have $p_{ij} = p_i \cdot c_{j\to i}$ if $i \neq j$. If $i=j$, we have $w_{ii} = (1-p_i) + p_i \cdot c_{j\to i}$ . In other words, if individual $i$ is socially influenced by any others, it implies $c_{i\to i}=0$, and we have $p_{ii} = (1-p_i)$ for any individual with input social connection from others. As for an isolated individual, it implies $p_{ii} = (1-p_i)+ p_i \cdot c_{ii} =1$ because we define $c_{i\to i}=1$ for any isolated individual as mentioned above. It is easy to prove that $\Sigma_j\, p_{ij}=1$, meaning that $P=[p_{ij}]_{n\times n}$ is a standard stochastic matrix, i.e., a non-negative matrix with all its rows summing up to 1.

Now given $N$ different individuals, it is natural to define a matrix to describe their relationship, and each element in the matrix is $p_{ij}$ , namely $P=[p_{ij}]_{n\times n}$. This is the first important matrix that defines relationship of individuals, and it could be time-varying. Here $p_i$ in our model usually changes slowly in normal cases because it indicates tendency of following one's own opinions or following others. This characteristics is inherently related to one's personality which is not easily changed. In contrast $c_{j\to i}$ could change much more often than $p_i$ as environmental surrounding changes, and it usually indicates whether one is talking to others or has any access to learn from others. For example, if you meet an acquainted person on the street and have a talk, it implies that $c_{j\to i}$ becomes non-zero such that you could exchange opinions with others. After the brief talk $c_{j\to i}$ returns to zero, implying that opinion impact from $j$ to $i$ is complete, and you continue to do other things. However, $p_i$ is still valued the same as before because your personality is not easily changed due to such a talk.

In sum the mathematical model given by Equation (4.3) could be well explained from the psychological perspective. Here parameter $p_i$ indicates one's inherent personality regarding stubbornness or persistence. Decreasing $p_i$ indicates one's tendency of holding his or her own attitude and judgment while increasing $p_i$ means inclination to accept others' opinions. Here $c_{j\to i}$ indicates whether one has access to learn from other opinion. In our particle-like model $c_{j\to i}$ is updated timely as individual move and selectively interact with others. In a general sense $c_{j\to i}$ represents a kind of social topology between individual $i$ and surrounding ones, and the entire social topology is summarized by a matrix $C=[\, c_{j\to i}\,]_{n\times n}$ . Usually perception and interaction among individuals are anisotropic, and thus $C$ is not symmetric commonly.

Furthermore, it is also feasible to split $p_{ij}$ as two components, namely $p_{ij} = p^{op}_{ij} + p^{be}_{ij}$, and $p^{op}_{ij}$ or $p^{be}_{ij}$ are the probability for individual $i$ to either learn from others' opinion directly or from others' behavior indirectly, namely $p^{op}_{ij} = p_i \cdot c_{j\to i} \cdot c^{op}_{j\to i}$ and $p^{be}_{ij} = p_i \cdot c_{j\to i} \cdot c^{be}_{j\to i}$ . Here $c^{op}_{j\to i}$ and $c^{be}_{j\to i}$ are conditional probability measure in a hierarchical structure such that $c^{op}_{j\to i} + c^{be}_{j\to i}=1$, indicating how much individual $i$ is impacted by individual $j$'s opinion or behavior. Thus, the whole picture of our opinion dynamical model is illustrated in Figure 4.2.

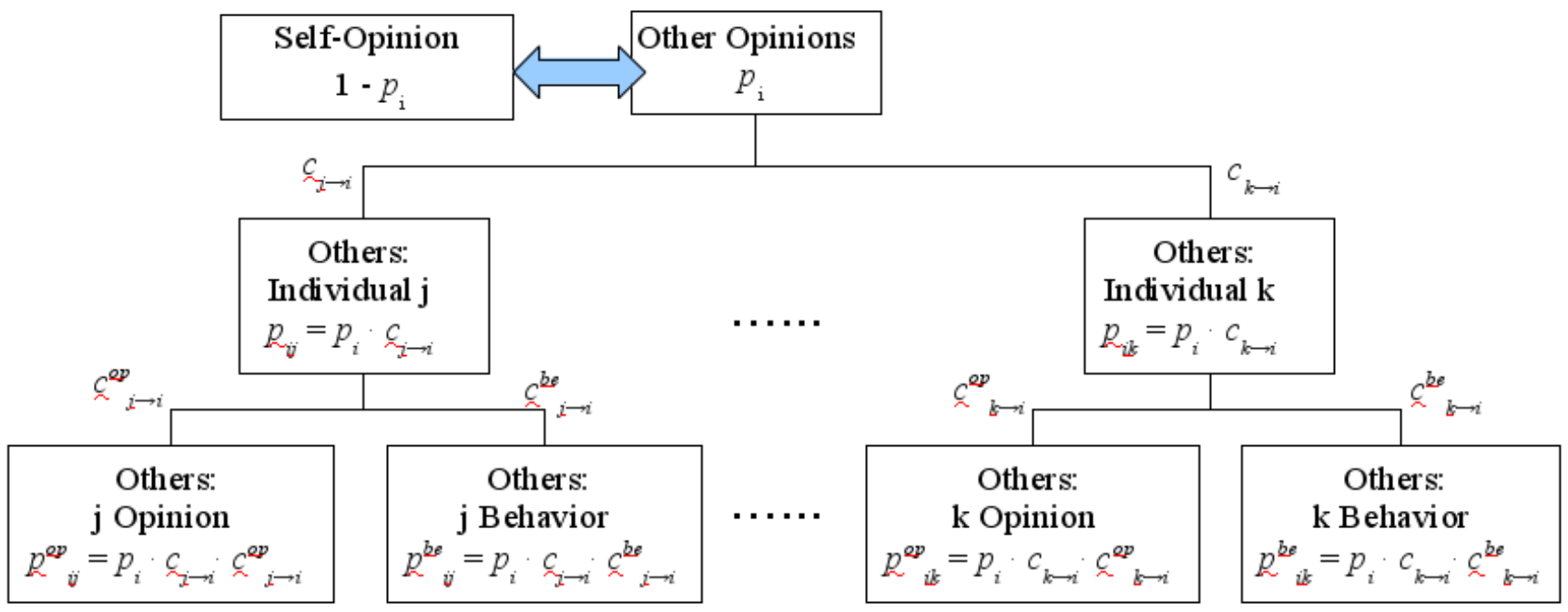

**Figure 4.2. Hierarchical structure of the opinion dynamics of individual i: The individual will balance between his or her own decision and others' decision, and others' decision are acquired from either observing their behavior or exchanging opinions by language.**

For simplicity, sometimes we may assume $c^{op}_{j\to i}$ and $c^{be}_{j\to i}$ are binary exclusive variables such that either $c^{op}_{j\to i}=1$ or $c^{be}_{j\to i}=1$, and the other one is zero. This simplified setting implies that individual *i* could follow either *j*'s opinion or behavior, but not mix them together. In the following discussion we focus on two-layer structure in above figure, and do not use $c^{op}_{j\to i}$ and $c^{be}_{j\to i}$ in our analysis. In other words we only care about whether individuals have access to acquire others' opinion, but do not further differ whether they acquire others' opinion by talking or observing others' behavior. However, in a psychological sense we emphasize that opinion and behavior are two different issues conceptually, and Figure 4.2 thus provides a general framework of modeling interactive opinion dynamics.

(b) Meaning of Parameter $c_{j\to i}$ and $p_i$

Equation (4.3) assumes that one's opinion is weighted with parameter (1-$p_i$) and $p_i$, and thus is updated by the weighted average. An important issue relates to the meaning of parameter $p_i$ and how it evolves in simulation. In a statistical sense $p_i$ is probability that individual expects to follow others. For a specific individual *j*, $p_{ij} = p_i \cdot c_{j\to i}$ measures how much individual *i* follows individual *j* in a probability sense, or in a reverse sense how much individual *j* could impact individual *i*'s opinion.

In Helbing, Farkas and Vicsek, 2000 and Helbing et al., 2002, $p_i$ is considered to indicate one's panic level, and it is given by ratio of $(v_i^0 - v_i)/v_i^0$, and it is called a "nervousness" parameter. This ratio critically affects several testing results in their work. As mentioned before the gap of $(v_i^0 - v_i)$ is understood as an indicator of one's stress level (See Table 3.1), and it is further normalized by dividing $(v_i^0 - v_i)$ by $v_i^0$. As a result, the "nervousness" parameter $(v_i^0 - v_i)/v_i^0$ can be explained as a normalized stress indicator, and it shows that people are more inclined to follow others when they feel more stressed in an emergency situation. This is a reasonable assumption and is consistent with psychological findings. An interesting issue in Helbing's model is that $v_i^0$ is used to adjust $p_i$, and $p_i$ is next used to calculate $v_i^0$. As a result, the information is looped in the model, and $p_i$ is not a stationary concept, but dynamically adaptive to the surroundings.

From the perspective of opinion dynamics the effective range of $c_{j\to i}$ is further extended as $c_{j\to i} \in [-1, 1]$ with $\sum_j |c_{j\to i}|=1$. As a result, $c_{j\to i}$ is not a probability measure, but a signed parameter which causes an individuals' opinion to either converge or diverge from another specific one. In brief, $0< c_{j\to i} \leq 1$ means that individual *i* intends to merge his or her opinion towards individual *j*. In contrast, $-1 \leq c_{j\to i}<0$ means that one is against individual *j*. In other words, the negative value of $c_{j\to i}$ implies disagreement such that individual *j*'s standpoint has an negative effect on individual *i*. As a result, the more individual *i* acquires opinion of individual *j*, the more he or she will reject it and hold more firmly on his or her own opinion. This brings the individuals' opinions not to converge, but diverge to some extent.

Table 4.2. Range of Parameter *p* in Opinion Dynamics

| $-1 \leq c_{j\to i}<0$ | $c_{j\to i} = 0$ | $0< c_{j\to i} \leq 1$ |
|---|---|---|
| Against others' opinions | Hold his own opinion and do not care about others' opinions | Support others' opinions |

In sum, $c_{j\to i} \in [-1, 1]$ implies a more complex social connection among individuals, where interactions bring opinions either closer to each other, or more apart from each other as shown in Table 4.2. Correspondingly, we should extend $\sum_j c_{j\to i} = 1$ to $\sum_j |c_{j\to i}|=1$. By extending the weight $c_{j\to i}$ from [0,1] to [-1,+1], we integrate a kind of "repulsion" among interacting opinions to describes both attraction and repulsion in a unified framework. This model is complex in theoretical analysis, but it is more comparable to real-world situations because people do not always agree with each other. The above model can be viewed as an extension of the signed networks, where $c_{j\to i} \in [-1, 1]$ implies that either agreement or disagreement among individuals (Altafini, 2013). As an extension of Equation (4.3), the above model can be practically computed by a asynchronous algorithm as below. Note that $\sum_j |c_{j\to i}|=1$ holds for j=1, 2, ...*n,* which forms a probability distribution for each agent. Thus, when randomly selecting an agent *j* with $c_{j\to i} \neq 0$ in the above algorithm, we can select one from the distribution $\sum_j |c_{j\to i}|=1$. If agent *i* is an isolated one, it gives that $c_{j\to i}=0$ for $\forall j \neq i$ and $c_{i\to i}=1$ such that the agent can only select himself or herself. Thus, Algorithm 1 is also reshaped as follows.

**Algorithm**: Pseudocode for asynchronous approach:

for each simulation time step *t* do

for each agent *i* in population do

select an agent *j* from distribution of $|c_{j\to i}|$, j=1, 2, ...n update opinion: $opinion_i(t+1) = (1- p_i \, \mathrm{sign}(c_{j\to i}))\, opinion_i(t) + p_i \, \mathrm{sign}(c_{j\to i})\, opinion_j(t)$

**end for**

**end for**

In Equation (4.1) and (4.2) the herding effect takes place only if the physical distance is less than $R_i$. In Equation (4.3) this implies that $c_{j\to i}$ becomes non-zero if $d_{ij} < R_i$, and a simple example is equipartition of weighting coefficient $c_{j\to i}$ for all the individuals in range of $R_i$. This is a reasonable and relatively straightforward assumption. According to various models in opinion dynamics, there are several models which are applicable to improve this assumption. For example, an existing theory suggest that interactions bring opinion closer to each other if they are already close sufficiently, and thus one's opinion is often inclined to selectively follow similar opinions of others. As a result, equipartition method will be replaced by weighted-average method such that the weighting coefficient $c_{j\to i}$ is larger for those with similar opinions. Moreover, the social relationship of individuals are also useful, and the interaction range is not only determined by physical distance $d_{ij}$, but also the desired distance $d_{ij}^0$. As a result, one tends to follow those in close social relationship.

Very interestingly, Equation (4.1) and (4.2) implies that one affects the surrounding people and is also affected by surroundings. Such interaction is mutual in nature, but it is not symmetric, and thus does not obey Newton 3rd Law.

Last but not least, whether Equation (4.1) and (4.2) will result in convergence of desired motion in a collective sense is another interesting topic to study. Current simulation results seem to be chaotic. However, existing psychological studies suggest that individuals' opinions may have the tendency of converging to a crowd opinion when they interact in certain circumstances. Suppose *N* individuals interacts, and their opinions are vectorized by *OPIN*(t) at time t, and it evolves by using the above model $OPIN(t+1)=P \cdot OPIN(t)$, where $P=[p_{ij}]_{nxn}$ is defined previously. Existing theory in linear algebra suggests that *OPIN*(t) could reach a stable solution if matrix *P* satisfies certain conditions. As a result, everyone's opinions (e.g, $\boldsymbol{v}_j^0(t)=v_i^0 \boldsymbol{e}^0_i$) may converge to a common value, and crowd opinion emerges. Extensive research is conducted in the field of opinion dynamics regarding this issue (Hegselmann and Krause, 2002).

(c) Opinions and Path-Selection Probability

We are surrounded by opinions, engaging with views on a diversity of issues such as economic and political processes, evolution of social norms. Such opinions are mathematically denoted by either continuous or discrete variables. In our model we may replace opinions by either desired velocity or desired interpersonal distance. In Equation (4.1) and (4.2) we give an example of replacing opinions by desired velocity. Another typical feature is the desired interpersonal distance $d_{ij}^0$.

In previous section we mentioned that $d_{ij}^0 \neq d_{ji}^0$, implying that the desired interpersonal distance is initially not balanced between two different individuals. However, in certain circumstances such as talking behavior, it is reasonable to assume that $d_{ij}^0$ and $d_{ji}^0$ converge to a same value such that two individuals become a stable pair in physical positions when interacting. Thus, the asynchronous algorithm can be used in pair as $d_{ij}^0(t+1)=(1-p_i)d_{ij}^0(t) + p_i d_{ji}^0(t)$ and $d_{ji}^0(t+1)=(1-p_j)d_{ji}^0(t) + p_j d_{ij}^0(t)$, and the algorithm assumes that individual *i* is talking to individual *j* such that $c_{j\to i}>0$ and $c_{i\to j}>0$. In Figure 4.3 we illustrate how $d_{ij}^0$ and $d_{ji}^0$ converge to the same value for two individuals. Besides, if $c_{j\to i}<0$ or $c_{i\to j}<0$, it is necessary to use the bounded confidence model to realize a stable system dynamics.

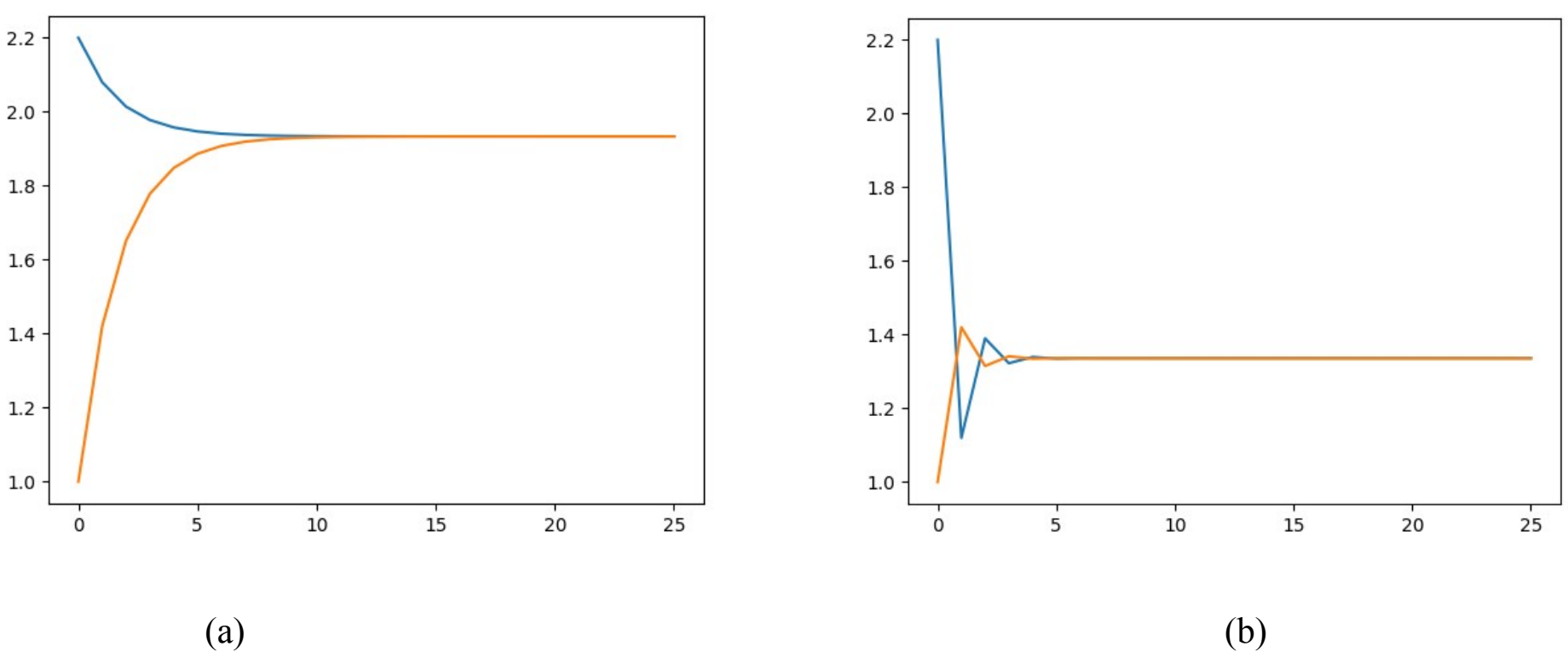

**Figure 4.3. Interaction of $d_{ij}^0$ and $d_{ji}^0$ by using opinion dynamics: $d_{ij}^0$ and $d_{ji}^0$ converge to a common value. The blue line indicates dynamics of individual *i* while the red line is for individual *j*. Initially $d_{ij}^0$ = 2.2 while $d_{ji}^0$ =1.0. In plot (a) $p_i$ = 0.1 and $p_j$ =0.35 while in plot (b) $p_i$ = 0.9 and $p_j$ =0.35.**

Furthermore, Equation (4.3) is a general form which can be applied to many features of individuals, such as path-selection probability or pre-evacuation time. By jointly using agent-based model with opinion dynamics, collective opinion formation is studied in social context. The problem is interesting because we discuss how individual decisions interact with each other in a time-varying social topology, and whether or in what conditions such interactions lead to a collective pattern or choice like consensus and polarization. As below we present an example where a group of individuals select an exit in a two-exit layout, and opinion is described by probability distribution of selecting the exits. The social typology is time-varying as individuals move and interact, implying that $P$=[$p_{ij}$ ]$_{nxn}$ is time-dependent. For simplicity we keep $p_i$ as a constant while $c_{j\rightarrow i}$ is timely updated as individuals move and perceive surrounding changes. The numerical testing results demonstrate consensus, polarization or clustering given different initial social typology conditions. The whole system is inherently nonlinear. Because the social topology is not stationary, but dynamically adaptive to the surrounding people, each individual's choice is selectively interacting with others' choices. The computational result is exhibited in Figure 4.4.

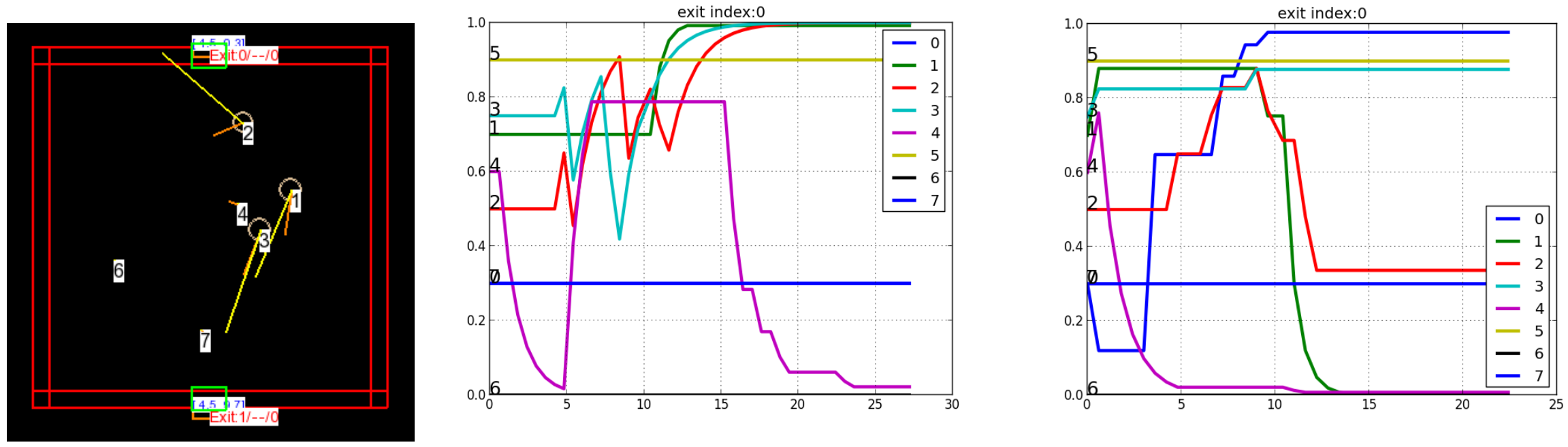

**Figure 4.4. Agents to select one door of two: Their opinions interact with each other and form group opinion. The above result shows dynamics of clustering based on social topology of individuals. The choice probability converges as shown above. However, even their opinions of probability distribution reach a consensus, for example, a distribution of [0.5, 0.5], their choices are statistically distributed evenly as half of them choose Exit 0 and the other half choose Exit 1. Such a consensus in opinions does not lead to a consensus in behavior, and this seems not well consistent with real-life scenarios, where individuals may abandon their individual choice in order to keep himself or herself in a group, as called "automaton conformity" in socio-psychological studies.**

## 4.2 Extension of Social Force

In a group individuals exhibit some degree of social cohesion based on their relationship and they are more than a simple collection or aggregate of individuals. To model group dynamics, attraction is necessarily taken into account in interaction of individuals. For example, attraction will make acquainted people to join together and possibly form a group. In Helbing and Molnar, 1995 and Helbing et al., 2002 attraction was considered, but separate from the social force. In this paper attraction and repulsion are put in the same social context: repulsion makes people to keep proper distance while attraction makes them cohesive and form groups. Thus, this subsection integrates attraction into the social force based on the concept of desired interpersonal distance. The resulting force is either repulsive or attractive, and it is especially useful to model group behavior in pedestrian crowd. The group social force is defined as below.

$$\boldsymbol{f}_{\boldsymbol{ij}}^{soc} = A_{ij}(d_{ij}^0 - d_{ij})\exp\left[\frac{(d_{ij}^0-d_{ij})}{B_{ij}}\right]\boldsymbol{n}_{\boldsymbol{ij}} \quad \text{or} \quad \boldsymbol{f}_{\boldsymbol{ij}}^{soc} = \left(\lambda_i + (1-\lambda_i)\frac{1+\cos\varphi_{ij}}{2}\right)A_{ij}(d_{ij}^0-d_{ij})\exp\left[\frac{(d_{ij}^0-d_{ij})}{B_{ij}}\right]\boldsymbol{n}_{\boldsymbol{ij}} \tag{4.4}$$

When $d_{ij}$ is sufficiently large, the group social force reduces to zero so that individual $i$ and $j$ have no interaction. This trend is the same as the repulsive social force given by Equation (2.3). If $d_{ij}$ is comparable to $d_{ij}^0$, interaction of individual $i$ and $j$ comes into existence. If $d_{ij}^0 < d_{ij}$, the group social force is attraction whereas it is repulsion if $d_{ij}^0 > d_{ij}$. The attraction reaches the extreme value when $d_{ij}^0 - d_{ij} = -B_{ij}$, where the extreme value is $\boldsymbol{f}_{ij}^{soc} = -A_{ij} B_{ij} \exp(-1)$. The desired distance $d_{ij}^0$ makes the curve move horizontally with a certain interval, and it is the equilibrium position when an individual interacts with another one in pair. The curve shape is affected by parameter $A_{ij}$ and $B_{ij}$. $A_{ij}$ is a linear scaling factor which affects the strength of the force whereas $B_{ij}$ determines the effective range of the interaction. In an addition, the force is approximated by a linear form when $d_{ij}^0 \approx d_{ij}$, and such a linear approximation will be useful in our later discussion of fluid dynamics for high-density crowd.

An advantage of the above group social force exists in characterizing the social relationship of individuals. Here two plots of Equation (4.4) are illustrated in comparison with another plot of traditional social force: Figure 4.5(a) shows that individual $i$ is attracted by individual $j$ when they stay close sufficiently, and individual $i$ is probably familiar with individual $j$. In contrast Figure 4.5(b) does not show such relationship because there is almost no attraction when they are close enough, and this plot is useful to describe interaction of strangers.

Moreover, the traditional formula of social force is compared with the group social force given by Equation (4.4), and it is noticed that the desired distance $d_{ij}^0$ is larger than $r_{ij}$, and parameter $A_i$ and $B_i$ are also in different values. In general, the traditional social force is usually considered as short-range interaction, and it plays a role of collision avoidance because it is calculated by using the physical size of individual agents (i.e., $r_{ij}$). In other words, the social force is effective only when people are very close to each other ($A_i$=2000; $B_i$=0.8), and it is fit to model high-density crowd. As for the new formula, it is relatively a long-range interaction where the desired distance $d_{ij}^0$ is commonly larger than $r_{ij}$, and parameter $B_{ij}$ of group social force is often larger than $B_i$ in the traditional formula, and parameter $A_{ij}$ of group social force is accordingly decreased such that $A_{ij} B_{ij}$ is within a reasonable range. In our numerical testing, it is found that $B_{ij}$ is usually in the range of $10^1 \sim 10^{-1}$ while $A_{ij}$ is commonly in the range of $10^2 \sim 10^0$. This issue will be further discussed in detail in numerical testing results.

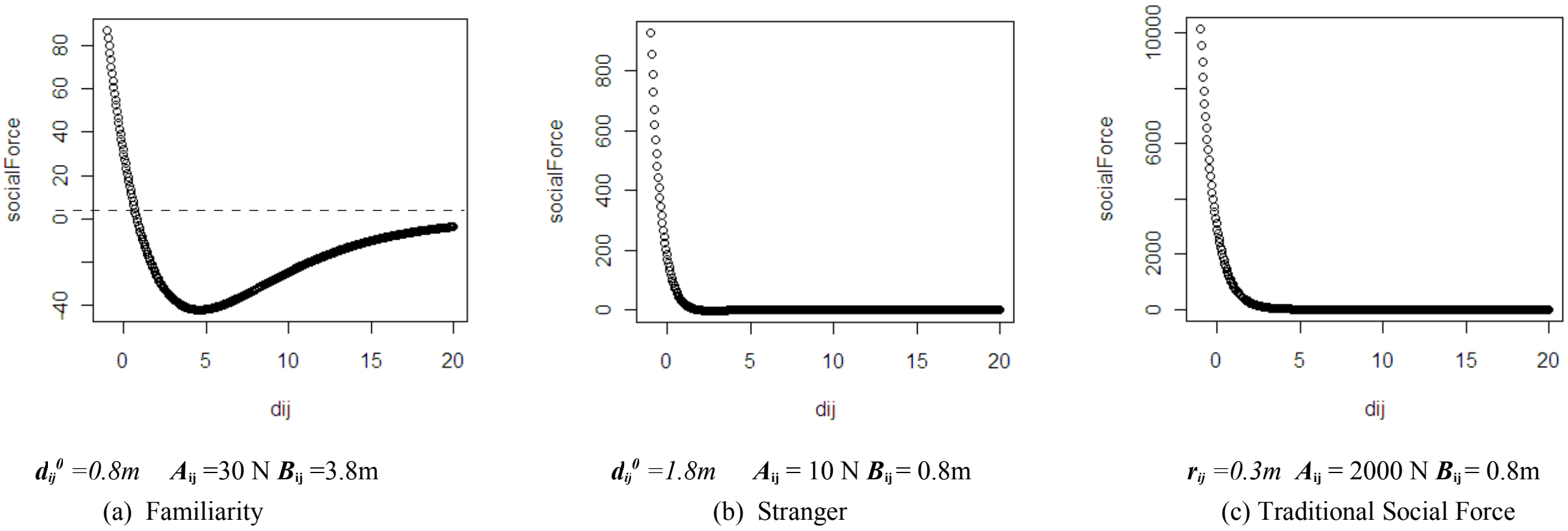

$d_{ij}^0$ =0.8m $A_{ij}$ =30 N $B_{ij}$ =3.8m
(a) Familiarity

$d_{ij}^0$ =1.8m $A_{ij}$ = 10 N $B_{ij}$= 0.8m
(b) Stranger

$r_{ij}$ =0.3m $A_{ij}$ = 2000 N $B_{ij}$= 0.8m
(c) Traditional Social Force

**Figure 4.5 Extended social force from individual *j* to individual *i (non-anisotropic formula)*: (a) To characterize two individuals who know each other, the force includes a negative segment representing attraction as well as a positive segment representing repulsion; (b) When two individuals are strangers, there is almost no attraction; (c) The traditional social force does not include attraction.**

In the above curve the negative segment represents attraction (See Equation 4.4), and it represents a kind of cohesive force. In contrast the positive segment denotes repulsion. When two individuals are strangers, there is almost no attraction as illustrated in Figure 4.5(b), and their interaction force is mainly repulsive. In order to differentiate the new formula from the traditional social force and highlight the cohesive feature in group dynamics, we will name Equation (4.4) by group social force or extended social force in this paper.

Equally importantly, the gap between $d_{ij}^0$ and $d_{ij}$ is expressed in Equation (4.4), and the interpersonal stress is characterized in consistency with our previous discussion. The gap of $d_{ij}^0$ and $d_{ij}$ is either negative or positive, meaning that being too far away or too close to someone result in stress in proximity. Keeping proper distance with others is the way to protect us from too much arousal, and this is evident in psychological study because being isolated or overcrowded can both lead to stressful conditions.

In addition, Equation (4.4) implies that $d_{ij}^0$ may be different from $d_{ji}^0$. As a result, the social-force between two individuals is not balanced, i.e., $d_{ij}^0 \neq d_{ji}^0$ and $\boldsymbol{f}_{ij}^{soc} \neq \boldsymbol{f}_{ji}^{soc}$. Thus, Newton third law does not hold for group social force either. Here an important issue is whether the social-force model means that pedestrian motion does not obey Newton 3rd law. The answer is definitely no because $d_{ij}^0$ is not a physics entity. In a psychological sense $d_{ij}^0$ is a subjective concept which exists in people's mind, and it characterizes how an individual intends to interact with others. As a result, the social-force given by Equation (2.3) and (4.4) and the self-driving force are both subjective forces which are generated involving one's intention and opinion. In a physics sense the subjective forces are generated by the foot-floor friction, which exactly obey Newton's laws. In other words Newton 3rd law is valid in pedestrian modeling at the physical level where the social force is viewed as a part of foot-floor friction. At another level where consciousness and opinions are critically involved to characterize how the social-force is generated, Newton's 3rd law is not applied. Therefore, the social-force model exhibits a bridge between the physics laws and psychological principles regarding crowd motion.

### 4.3 Grouping Dynamics

People create group-level behavior beyond the ken of any single person, and in the past 20 years there has been growing realization in social science that such group-level organizations sometimes emerge spontaneously without any central design. Thus, it is reasonable to study such group phenomena in a bottom-up rather than a top-down manner. This approach creates computational units of individuals and their interactions, and to observe how the global structures are formed dynamically with their interactions.

Very importantly, we think both consciousness and unconsciousness contribute to form group-level behavior. This effect was initially described in LeBon's famous book, "The Crowd: A Study of the Popular Mind"(LeBon, 1895). In this book the author described a phenomenon that an individual seems forgetting the original motive when immersing oneself in the crowd, and thus follow the collective motive of the crowd. This process involves deindividualization with unconsciousness, and we will also try to model this process in our bottom-up modeling framework.

With combination of social cohesion and herding behavior, a kind of convergent pattern is supposed to emerge in a crowd. Here the social groups and herding effect are related but different concepts. Social groups describe the social relationship of individuals, and it emphasizes whether there is a social tie between individuals, and such a social tie facilitates to form a group. The herding effect, or generally considered as opinion dynamics, emphasizes how an individual's opinion interacts with others' to form a common motive or destination in collective motion. You may meet your friend on the street, but if you do not have a common destination, you and your friend usually head to each destination individually after greeting or talking briefly. Another example is evacuation of a stadium where people follow the crowd flow to move to an exit. There are a multitude of small groups composed of friends or family members, and they keep together in egress because of their social relationship. These small groups also compose a large group of evacuees, and herding behavior widely exists among these small groups, contributing to form a collective pattern of motion. In sum, the group social force makes individuals socially bonded with each other, and it emphasizes the social relationship of individuals. Herding effect does not focus on such social relationship, but emphasizes people tend to follow their neighbors' characteristic, and thus help to form a common motive.

Considering a group composed by $n$ individuals, the social relationship of the group members is described by a $n$x$n$ matrix $D^0$, of which the element is $d_{ij}^0$. In a similar way, there are $n$x$n$ matrices $A$ and $B$, and the elements are $A_{ij}$ and $B_{ij}$, respectively. Generally speaking, $D^0$, $A$ and $B$ are asymmetrical.

$$\boldsymbol{A}=[A_{ij}]_{n\times n} \qquad \boldsymbol{B}=[B_{ij}]_{n\times n} \qquad \boldsymbol{D^0}=[d_{ij}^0]_{n\times n} \tag{4.5}$$

The extended social force is specified by the matrices $D^0$, $A$ and $B$, and the method has been tested in FDS+Evac as well as another egress simulator we have developed. A testing result is illustrated in Figure 4.6, where the built environment is set up based on an example in Pan et. al., 2007. Several groups are identified in the simulation. Some small groups merge into a large group and regrouping may occur at intersections or at bottlenecks when many groups meet there. In sum, grouping behavior is not a static concept in our model, but an adaptive feature. Structure of groups change dynamically, resulting in a self-organized phenomenon during the movement.

Based on Equation (2.2) and (4.4), a typical kind of social relationship is described as the leader-and-follower group, where there is a kind of individual whose behavior is mainly motivated by himself, and if others would like to follow them, they become leaders in a group. Thus, if individual *i* is the leader in a group, his motion is mainly motivated by the self-driving force (i.e., desired velocity). Follower is the one whose behavior is mainly motivated by others. Thus, if individual *i* is the follower in a group, his motion is mainly governed by group social force, and the self-driving force is secondary. As mentioned before imbalance (asymmetry) of $d_{ij}^0$ and $d_{ji}^0$ will contribute to model leadership in crowd behavior. If *i* is a leader, $d_{ji}^0$ is much smaller than $d_{ij}^0$. As a result, the leader will attract his surrounding people, but not easily be attracted by them. In brief an individual's motion can be classified into two types. One type of motion is primarily motivated by the self-driving force and thus is called active motion. The other type of motion is largely motivated by surrounding people, and is called passive motion. In general, an individual's motion is a combination of both types, but we can differentiate such two types in simulation and identify whether one's motion is dominated by either active or passive type.

As shown in Figure 4.6 we use the color bar to observe the magnitude of the group social force in simulation (Korhonen 2017; Forney, 2018; McGrattan et al., 2018). An individual in active motion often moves in the front of a group. Individuals in passive motion are followers in the group and they usually move behind the leader. The leader is commonly under smaller group social force than the followers.

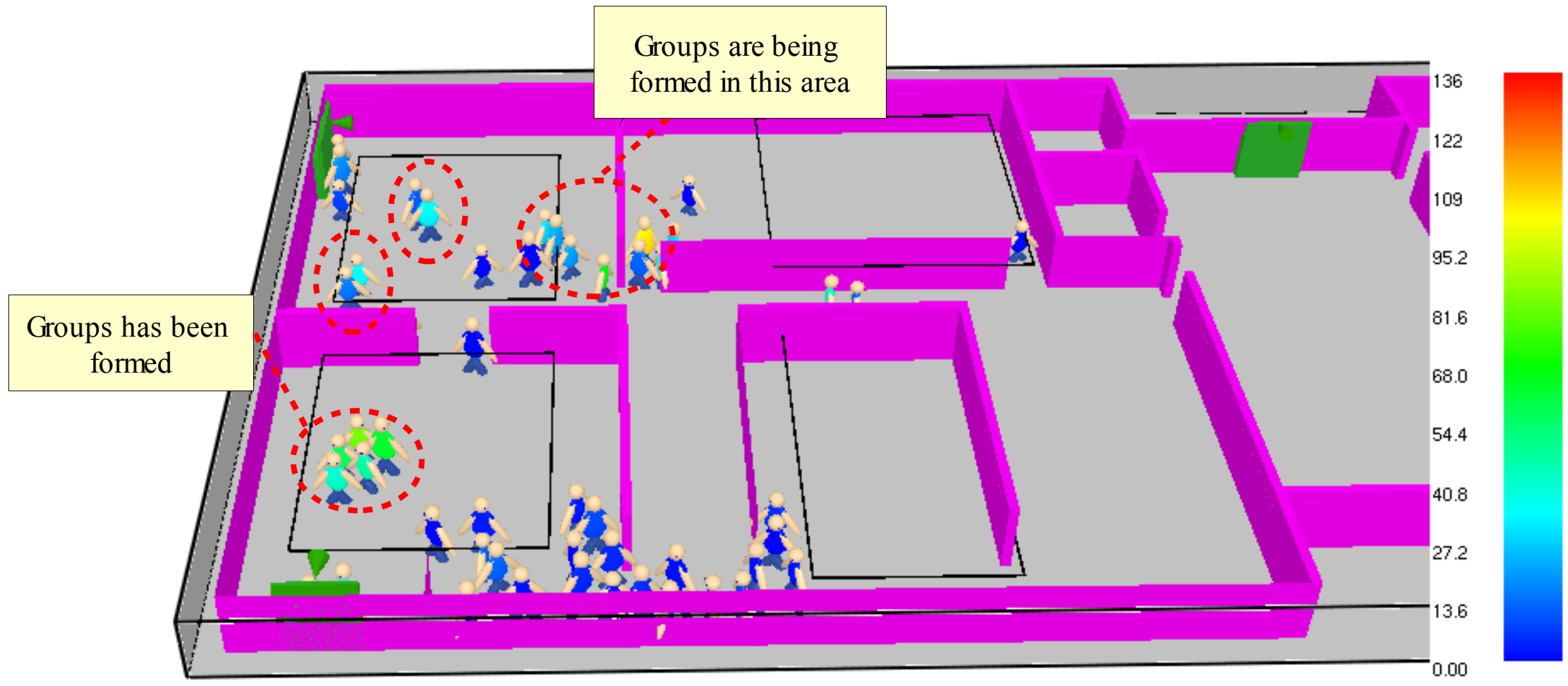

**Figure 4.6. Simulation of Group Dynamics in FDS+Evac: The color bar indicates the magnitude of the extended social force given by Equation (4.4). An individual in active motion usually moves in the front of a group. Individuals in passive motion are followers in the group and they usually move behind the leaders. The leaders are under smaller extended social force than the followers.**

About group dynamics we have the following two remarks regarding social and psychological studies.

In evacuation simulation Equation (4.4) and (4.5) can be better explained by flight-or-affiliation effect in psychological studies. The self-driving force motivates one to flee while the group social force makes one affiliated with others. This effect agrees with social attachment theory in psychological study (Mawson, 2007; Bañgate et al., 2017). The social attachment theory suggests that people usually seek for familiar ones (e.g., friends or parents) to relieve stress in face of danger, and this is rooted from our instinctive response to danger in childhood when a child seek for the parents for shelter. Affiliated with familiar and trust individuals relieves our stress. Thus, different from the fight-or-flight response (Cannon, 1932), the modified social model well agrees with the flight-or-affiliation effect. Therefore, the pedestrian model presented above is especially useful to model crowd behavior in pre-movement stage in crowd evacuation (Sorensen, 1991, Kuligowski, 2009). In brief, when the alarm or hazard is detected, people usually do not head to exits immediately, but go to find their friends or trust ones to form groups. Such grouping effect usually delays the movement towards exits, and thus is called pre-movement stage. Thus, the new model contributes to modeling the crowd behavior in pre-movement period and will be useful to investigate how the initial delay is formed and influenced by the group dynamics.

In addition, can we describe collective unconsciousness based on our model? In fact, the diagonal elements in matrix *D*, A and B imply a kind of force to oneself, where $d_{ii} = 0$, but Equation (4.4) implies this force is zero because $\boldsymbol{n}_{ii}$ is a zero vector. However, we may modify the vector slightly such that an individual implements a kind of force to oneself. In Equation (4.6) $\boldsymbol{n}_{ii}$ is replaced by the opposite direction of the driving force $\boldsymbol{f}_i^{drv}$, and this force is called self-repulsion in this paper.

$$\boldsymbol{f}_{ii}^{soc} = A_{ii}(d_{ii}^0 - d_{ii})\exp\left[\frac{d_{ii}^0 - d_{ii}}{B_{ii}}\right]\boldsymbol{n}_{ii} = A_{ii} d_{ii}^0 \exp\left[\frac{d_{ii}^0}{B_{ii}}\right]\boldsymbol{n}_{ii} \quad \rightarrow \quad \boldsymbol{f}_{ii}^{soc} = A_{ii} d_{ii}^0 \exp\left[\frac{d_{ii}^0}{B_{ii}}\right](-norm(\boldsymbol{f}_i^{drv})) \tag{4.6}$$

Very interestingly, the self-driving force is understood as generated by conscious mind of an individual, and results in one's motivation of behavior. The self-repulsion refers to the unconscious mind, and it may be against the conscious motive that we are aware of. Thus, we assume that the direction of self-repulsion is contrary to the driving force which represents an individual's motive. This model is useful to characterize certain crowd behavior depicted in LeBon, 1895. That is, when an individual immerses oneself in the crowd, he or she may lose part of his individual feature such as his original motivation, and thus simply follow the collective motive of the crowd. Thus, the force specified by Equation (4.6) neutralizes the effect of self-driving force. Instead, the group social force will become dominant such that an individual's motive is replaced by the crowd motive, especially by the leader's motive in the crowd. In order to further describe this effect we may specify the self-repulsion by $\boldsymbol{f}_{ii}^{soc}=[1-\exp(-\beta)](-\boldsymbol{f}_i^{drv})$, and parameter $\beta \geq 0$ indicates level of one's immersion in the crowd, and it is calculated from elements in $D^0$, $A$ and $B$. In a sense $\beta=0$ denotes $\boldsymbol{f}_{ii}^{soc}=0$ such that one's conscious mind is independent, not influenced by other people. As $\beta$ increases, the self-repulsion $\boldsymbol{f}_{ii}^{soc}$ goes towards $-\boldsymbol{f}_i^{drv}$ so that the individualistic feature is neutralized by $\boldsymbol{f}_{ii}^{soc}$, meaning that people immerse themselves in the crowd. In sum subconsciousness is an interesting topic in psychological studies, and it is a fantastic issue to be further investigated.

## V. Oscillating Walkers in Collective Behavior

In the previous sections microscopic models are applied in 2-dimensional space, where individual pedestrian's motion is governed by a mathematical equation, and we simulate interaction of many individuals and observe collective patterns in the numerical results. This section will further study the mathematical equation of social-force model from the perspective of ordinary differential equation (ODE). The idea is to explain a confusion and controversy on oscillating phenomenon when an individual is approaching another one. This phenomenon is partly due to the dilemma of choosing proper parameters to avoid both overlapping and oscillations of the moving pedestrians (Chraibi, et al., 2011, Kretz, 2015).

More importantly, it has been clarified that the social force model agrees with Newton Laws in Section 3, and it is thus on a solid foundation of classic physics. The simulation result of the model is expected to be consistent with what we observe in reality. However, oscillating walkers are not observed in the realistic world. From the perspective of physics these problems are difficult to be addressed, and the problem is thus reviewed using a psychological perspective. The key issue to understand this problem is that the desired velocity and desired distance are not physics entities.

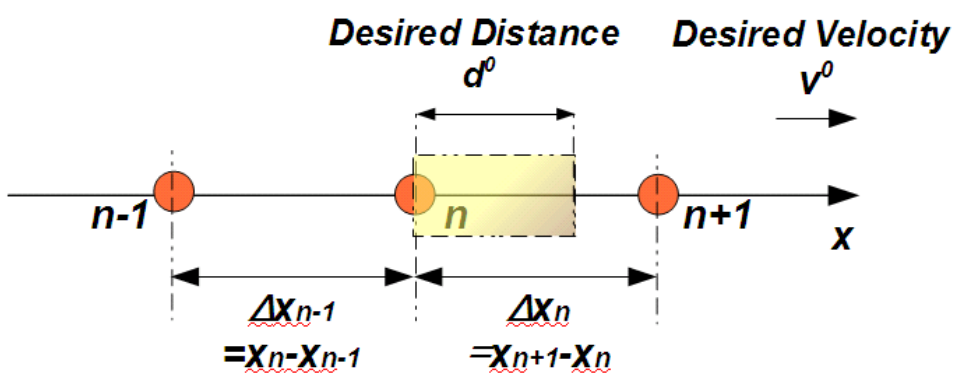

**Figure 5.1 Oscillating Agents in One-Dimensional Space: Three pedestrian agents are distributed uniformly as above. The physical size of agents is ignored for simplicity and the desired interpersonal distance $d^0$ is highlighted. Moreover, for interactions among agents, it is assumed that pedestrians have foresight such that pedestrian *n* is only influenced by the pedestrian in front, not by the one behind.**

In general pedestrian motion is commonly modeled in two-dimensional space. In this section we will simplify the scenario to one-dimension space to highlight oscillation phenomenon. In such a uni-directional space we can focus on the magnitude of velocity and forces, and their directions are integrated into algebraic values. Given $N$ pedestrians distributed uniformly in a corridor with closed boundary condition, each one is modeled as a simple geometric object of constant size. For simplicity the body size of individual pedestrian is omitted and the desired interpersonal distance $d^0$ is highlighted. Moreover, for interactions among $N$ pedestrians, we assume that pedestrians have foresight such that pedestrian $n$ is only influenced by the adjacent one in front, not by the one in behind. The location, velocity and acceleration of individual $n$ is hereby denoted by $x_n$, $dx_n/dt$ and $d^2x_n/dt^2$. The distance is denoted $\Delta x_n = x_{n+1} - x_n$, and we have the following equation by Newton Second Law.

$$\text{Equation for Newton Second Law: } m_0 \frac{d^2 x_n}{dt^2} = f^{drv} - f^{soc} \qquad (5.1)$$

| Interaction Force: Repulsive Social Force (Eq. 3.1) | $m_0 \frac{d^2 x_n}{dt^2} = \frac{m_0}{\tau}(v^0 - v) - A\exp\left(\frac{d^0 - \Delta x_n}{B}\right)$ |
|---|---|
| Interaction Force: Group Social Force (Eq. 4.4) | $m_0 \frac{d^2 x_n}{dt^2} = \frac{m_0}{\tau}(v^0 - v) - A(d^0 - \Delta x_n)\exp\left(\frac{d^0 - \Delta x_n}{B}\right)$ |

The above model is more or less similar to the famous Car-Following models (Bando, 1995; Chandler et al., 1958; Kachroo and Ozbay, 1999), where the car drivers adjust acceleration according to the conditions in front, and each vehicle is governed by an ordinary differential equation (ODE) that depends on speed and distance of the car in front. However, we highlight the psychological features for pedestrian motion, and will emphasize desired velocity and desired distance in the above equation, and the physical force is not taken into account.

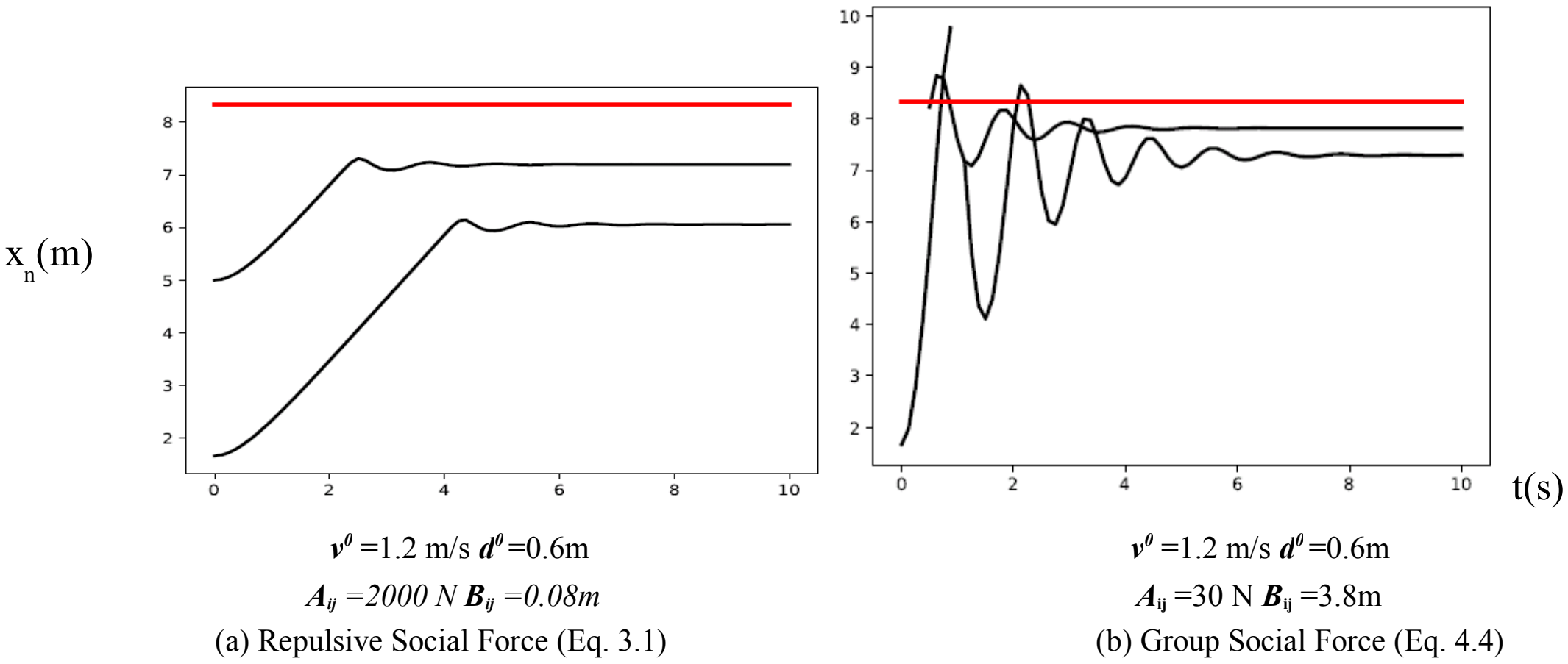

$v^0$ =1.2 m/s $d^0$=0.6m
$A_{ij}$ =2000 N $B_{ij}$ =0.08m
(a) Repulsive Social Force (Eq. 3.1)

$v^0$ =1.2 m/s $d^0$=0.6m
$A_{ij}$ =30 N $B_{ij}$ =3.8m
(b) Group Social Force (Eq. 4.4)

**Figure 5.2 Simulation of oscillating agents in 1D space: Both the repulsive social force and group social force result in oscillation phenomena. The oscillation is especially intensive when group social force is applied.**

In a system composed of $N$ pedestrians, $N$ equations are formulated, and it is feasible to simulate the system by a computer program. Now we assume a simple example where only two agents are simulated, and their desired velocity is given by $v_0$ =1.2m/s, and desired interpersonal distance is specified $d_0$ =0.6m. The simulation result is shown in Figure 5.2. Because physical interaction is not considered in Equation (5.1), one agent may surpass another when oscillation become intensive.

It is obvious that oscillation occurs when one gets close enough to another. The repulsive social force generate moderate oscillation, and the group social force causes intensive oscillation. However, such oscillations is not a phenomenon observed in reality. When a human gets close to another one in the realistic world, we never oscillate. In a sense we may oscillate mentally by feeling stress, but we do not oscillate in the physical position. Thus, a key question is how we will understand such oscillation and if there is any method to offset oscillation in our model and simulation.

## 5.1 Offset Oscillation By Control Theory

Technically, there are several existing methods to study oscillation. A common method is to linearize the motion equation at its equilibrium point and discuss its characteristic equation and characteristic root. Take Equation (5.1) for example. If the repulsive social force is applied, existing research justifies oscillation exists if $4\tau v^0<B$ (Kretz, 2015). Very interestingly, this result shows that oscillation is only related to the parameter of $\tau$, $v^0$ and $B$, and it does not depend on $d^0$ and $A$. Here we omit the physical size of agents in the above analysis.

How to offset oscillation is another important issue, especially for group social force. In fact the oscillation phenomenon is common in dynamic systems. From the perspective of classical control theory, there are a number of methods to offset the oscillation (e.g., PID controller). A widely-used method is adding a derivative term, namely, adding the relative velocity $v_{ij} = v_i - v_j$ as a force component to Equation (5.1). In one-dimensional space $v_{ij}$ is simplified by $-d\Delta x_n/dt = dx_n/dt - dx_{n+1}/dt$. The relative velocity of two adjacent individuals is an estimate of the future trend of their relative position $\Delta x_n = x_{n+1} - x_n$ based on its current rate of change. This derivative element in an important part in widely-used PID controller, and it is sometimes called "anticipatory control," which contributes greatly to reducing the oscillation in a dynamic system. This term has been traditionally applied in car-following models, where a car accelerates based on the relative speed of itself and the car in front. This feature has been used in several pedestrian models such as the generalized centrifugal force model (Chraibi et. al., 2011).

By using a force component as a function of relative velocity $v_{ij}$, the oscillation phenomenon will be significantly mitigated as shown in Figure 5.3. The simulation result strongly justifies necessity to use relative velocity $v_{ij}$ as a force component of group social force to avoid intensive oscillation in certain conditions. The force component is formulated in Equation (5.2) as a function of $v_{ij} = dx_n/dt - dx_{n+1}/dt$, where an individual will try to reduce the movement speed when approaching the one in front, and the stopping trend increases with the relative speed $dx_n/dt - dx_{n+1}/dt$ if it is positive, namely the one in front is approaching near, not far away. In addition, the stopping force increases with the relative distance $\Delta x_n = x_{n+1} - x_n$ to avoid any physical contact. This is basically from vehicle traffic problem where collision is definitely to be prevented.

$$m_0 \frac{d^2 x_n}{dt^2} = \frac{m_0}{\tau}(v^0 - v) - A(d^0 - \Delta x_n)\exp\left(\frac{d^0 - \Delta x_n}{B}\right) - \frac{c}{\Delta x_n} max\left(\frac{dx_n}{dt} - \frac{dx_{n+1}}{dt}, 0\right) \quad max(x, 0) = \begin{cases} x & \text{if } x \geq 0 \\ 0 & \text{if } x < 0 \end{cases} \qquad (5.2)$$

$x_n$(m)

t(s)

$\boldsymbol{v^0}$ =1.2 m/s $\boldsymbol{d^0}$=0.6m $\boldsymbol{\tau}$=0.5s

$\boldsymbol{A_{ij}}$ **/m₀** =20 N $\boldsymbol{B_{ij}}$ =3.6m

(a) Group Social Force (Eq. 4.4)

$\boldsymbol{v^0}$ =1.2 m/s $\boldsymbol{d^0}$=0.6m $\boldsymbol{\tau}$=0.5s

$\boldsymbol{A_{ij/}}$ $\boldsymbol{/m_0}$ =20 N $\boldsymbol{B_{ij}}$ =3.6m**Beta** $\boldsymbol{/m_0}$ =3.0

(b) Group Social Force with $\boldsymbol{V}$ij Component (Eq. 5.2)

**Figure 5.3 Two agents are simulated in 1D space: The simulation result strongly justifies it necessary to use relative velocity $\boldsymbol{v_{ij}}$ as a force component when the group social force is applied. Otherwise the group social force may generate intensive oscillation in certain conditions.**

## 5.2 Varying Opinions in Desired Velocity and Desired Interpersonal Distance

As above we discuss the oscillating walkers by control theory and highlight the function of $\boldsymbol{v}_{ij}$ to offset oscillation. There is another perspective of pedestrian modeling. In real-world scenario people usually have a target interpersonal distance and get there without oscillation in physical positions. We note that a major problem is that $\boldsymbol{v}^0$ and $d^0$ are assumed to be constant in the above equations. As mentioned before $\boldsymbol{v}^0$ and $d^0$ are not physical entities, but reflect people's opinions as shown in Table 3.1. If $\boldsymbol{v}^0$ and $d^0$ are not constant, pedestrians may adapt $\boldsymbol{v}^0$ and $d^0$ based on the physical velocity $\boldsymbol{v}$ and distance $d = \Delta x_n$ . In certain conditions it may be reasonable to infer that $\boldsymbol{v}^0$ and $d^0$ may oscillate instead of $\boldsymbol{v}$ and $d$ .

Recall the doorway example in Figure 3.1 and 3.2 where a number of individuals pass through a bottleneck, and we see that $\boldsymbol{v}^0$ and $d^0$ are not only time-varying, but also inter-related variables. In specific sense of emergency initially motivates people to move fast so that desired velocity $\boldsymbol{v}^0$ raises as initial cause of motion. At bottlenecks such motion cannot be realized as desired, and the gap of $\boldsymbol{v}^0 - \boldsymbol{v}$ increases. Such time-related stress is transferred to space-related stress as shown in Figure 5.4. The desired interpersonal distance $d^0$ at bottlenecks is thus reduced, and people would like to "compress" their personal space in order to pass through the bottleneck quickly. This effect actually exhibits a kind of collective intelligence to increase transport efficiency at the bottleneck. The social norm is thus modified such that $d_{ij}^0$ is scaled down at bottlenecks.

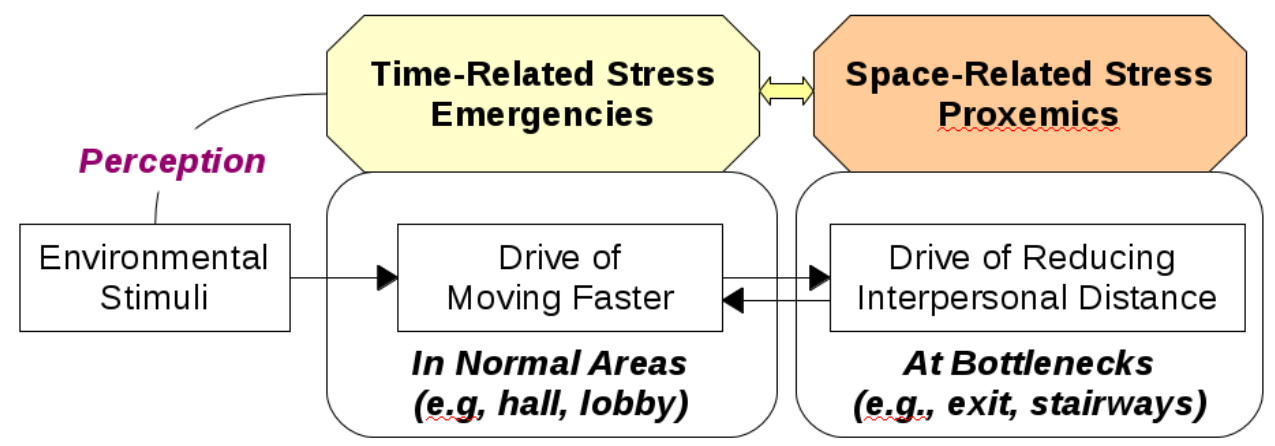

**Figure 5.4. Two Types of Stress Transforming Mutually: The sense of emergencies creates a kind of time-related stress which drives people to move fast in escape. At bottlenecks (e.g., narrow doorways) people cannot speed up as desired, and thus time-related stress is transformed into space-related interpersonal stress in order to pass through the bottleneck quickly.**

In the above analysis we understand that $\boldsymbol{v}^0$ and $d^0$ are inter-related. In principles $\boldsymbol{v}^0$ is time-based concept, and it is the fundamental cause of motion in most situations. In one-dimensional space we could formalize the above effect by formula $d_{ij}^0 = d_{ij}^{0\text{-preset}} \max(0.2, |\boldsymbol{v}_i|/|\boldsymbol{v}_i^0|)$, where $d_{ij}^{0\text{-preset}}$ is the desired interpersonal distance in normal situation and 0.2 is the lower bound factor. As a result, $d_{ij}^0$ is a function of $|\boldsymbol{v}_i|/|\boldsymbol{v}_i^0|$, and it decreases if $|\boldsymbol{v}_i|<|\boldsymbol{v}_i^0|$, implying that $\boldsymbol{v}_i^0$ also impacts the desired distance $d_{ij}^0$ and the social force. As a result, the social force decreases at bottlenecks. The tradeoff of reducing $d_{ij}^0$ at bottlenecks is increase of possibility of physical interaction among people, which is the major cause of crowd disasters such as stampede.

In the above analysis $d^0$ is a function of $v^0$. In certain situation such relationship could be reversed such that $v^0$ becomes a function of $d^0$. Consider a scenario where an individual is approaching another. When their physical distance $d$ is close to $d^0$, at least one of them will decelerate such that $v^0$ is expected to be decreasing to zero. In this process an individual adapts $v^0$ to desired distance $d^0$, and $\boldsymbol{v}^0$ is thus a function of $d^0$. A dynamic model of this scenario is given in Equation (5.3) where desired velocity $v^0$ decreases when individuals move close enough, namely $|d^0 - \Delta x_n| < R$. In this model $v_{ij} = -d\Delta x_n/dt$ is integrated as a part of desired velocity $v^0$, and it also plays an important role of reducing oscillation.

$$\frac{d^2 x_n}{dt^2} = \frac{(v^0 - v)}{\tau} - \frac{A}{m_0}(d^0 - \Delta x_n)\exp\left(\frac{d^0 - \Delta x_n}{B}\right)$$

$$|d^0 - \Delta x_n| < R: \; v^0 = c_1(d^0 - \Delta x_n) - c_2 \frac{d\Delta x_n}{dt} \qquad c_1<0 \text{ and } c_2<0;$$

$$|d^0 - \Delta x_n| \geq R: \; v^0 \text{ and } d^0 \text{ are determined by high-level algorithm such as door selection} \tag{5.3}$$

In the Equation (5.3) if people get close to the desired distance $d^0$, $v^0$ will reduce to zero accordingly. The interacting range $R$ in Equation (5.3) is a threshold that determines when the desired velocity becomes a function of $d^0$. In principle the model suggests that $v^0$ and $d^0$ are not independent with each other, but inter-related variables in people's mind.

The system given by Equation (5.3) is formally studied by its characteristic equation. Based on Equation (5.3) when agents are close enough (i.e., $|d^0 - \Delta x_n| < R$), the equilibrium point is at $dx_n/dt = 0$ and $d^2x_n/dt^2 = 0$, namely the velocity and acceleration are both zeros, and it gives $\Delta x_n = d^0$ and $v^0 = 0$, which means in the equilibrium position people tend to decelerate to $v=0$ such that they stand still and physical distance $\Delta x_n = d^0$ is realized. When $\Delta x_n$ is close to $d^0$, we linearize the group social force as below.

$$\frac{d^2 x_n}{dt^2} = \tau^{-1}\left(v^0 - \frac{dx_n}{dt}\right) - \frac{A}{m_0}(d^0 - \Delta x_n) \tag{5.4}$$

In order to simplify the mathematical notations we let $k_1 = 1/\tau$ and $k_2 = A/m_0$

$$\frac{d^2 x_n}{dt^2} + k_1 \frac{dx_n}{dt} - k_2 \Delta x_n = k_1 v^0 - k_2 d^0 \tag{5.5}$$

$$\frac{d^2 x_n}{dt^2} + k_1 \frac{dx_n}{dt} - k_2 (x_{n+1} - x_n) = k_1 v^0 - k_2 d^0 \tag{5.6}$$

Again we have $\Delta x_n = x_{n+1} - x_n$, and we assume that $x_{n+1}$ is a constant because $n$+1 agent stands still

$$\frac{d^2 x_n}{dt^2} + k_1 \frac{dx_n}{dt} + k_2 x_n = k_2 x_{n+1} + k_1 v^0 - k_2 d^0 \tag{5.7}$$

Now $v^0$ is plugged into the above equation as a variable, and it gives

$$\frac{d^2 x_n}{dt^2} + k_1 \frac{dx_n}{dt} + k_2 x_n = k_2 x_{n+1} + k_1 c_1 (d^0 - \Delta x_n) - k_1 c_2 \frac{d \Delta x_n}{dt} - k_2 d^0 \tag{5.8}$$

$$\frac{d^2 x_n}{dt^2} + k_1 \frac{dx_n}{dt} + k_2 x_n = k_2 x_{n+1} + k_1 c_1 (d^0 - x_{n+1} + x_n) - k_1 c_2 \frac{d (x_{n+1} - x_n)}{dt} - k_2 d^0 \tag{5.9}$$

$$\frac{d^2 x_n}{dt^2} + (k_1 - k_1 c_2) \frac{dx_n}{dt} + (k_2 - k_1 c_1) x_n = k_2 x_{n+1} + k_1 c_1 (d^0 - x_{n+1}) - k_1 c_2 \frac{d x_{n+1}}{dt} - k_2 d^0 \tag{5.10}$$

The characteristic equation is thus given as below.

$$\lambda^2 + (k_1 - k_1 c_2) \lambda + (k_2 - k_1 c_1) = 0 \tag{5.11}$$

The system described as above does not oscillate if the imaginary part of the solutions are null, i.e. if $k_1^2(1- c_2)^2 >= 4(k_2 - k_1 c_2)$, namely, $(1- c_2)^2/\ ^2 >= 4(A/m_0 - c_2/\ )$. Here we note that solution does not depend explicitly on $v^0$ and $d^0$ , but relates to the parameter , $A$ and $m_0$ in the model, where $\tau k_1 = 1$ and $A/m_0 = k_2$ .

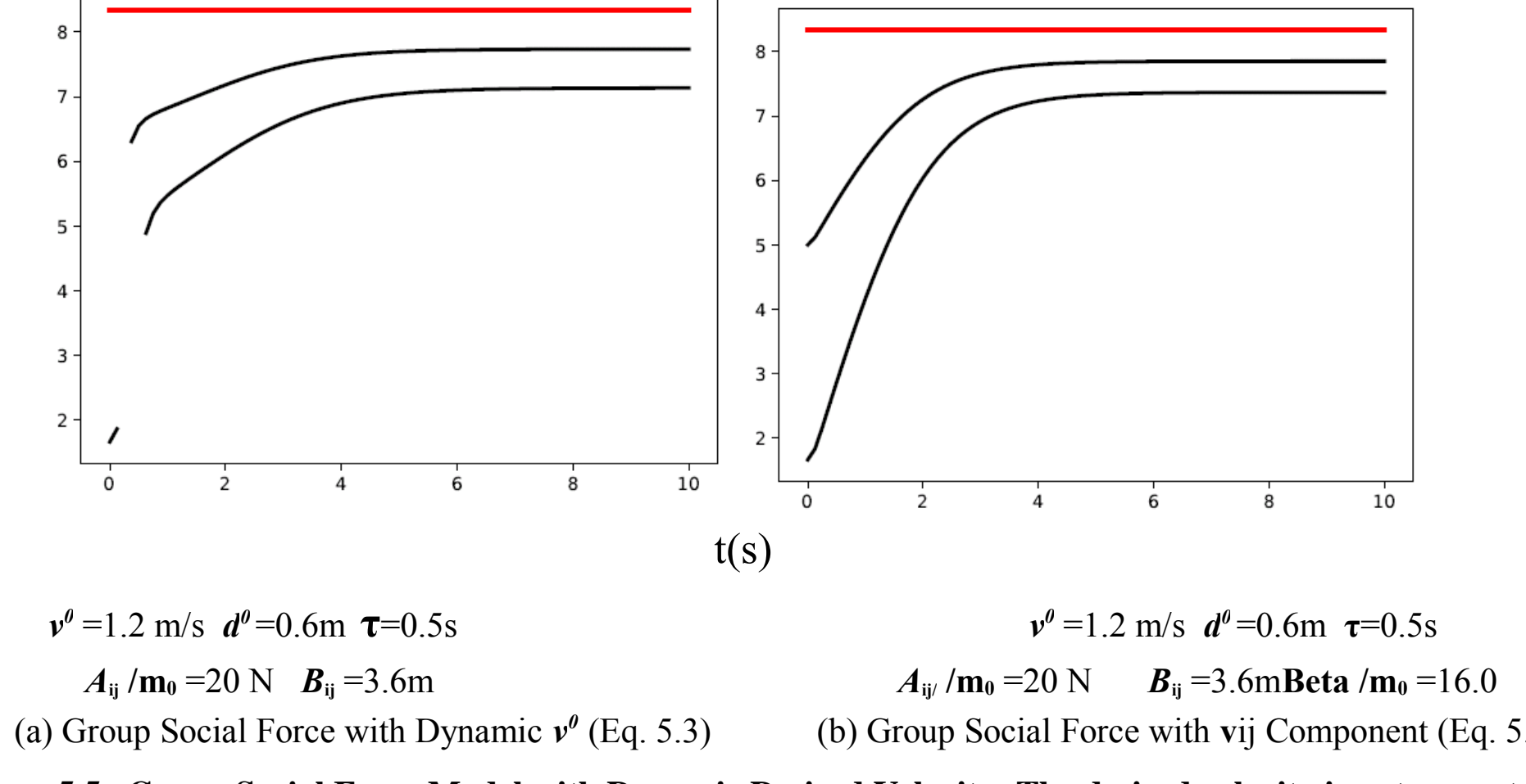

$v^0$ =1.2 m/s $d^0$=0.6m **τ**=0.5s

$A_{ij}$ /**m₀** =20 N $B_{ij}$ =3.6m

(a) Group Social Force with Dynamic $v^0$ (Eq. 5.3)

$v^0$ =1.2 m/s $d^0$=0.6m **τ**=0.5s

$A_{ij/}$ /**m₀** =20 N $B_{ij}$ =3.6m**Beta /m₀** =16.0

(b) Group Social Force with **v**ij Component (Eq. 5.2)

**Figure 5.5. Group Social Force Model with Dynamic Desired Velocity: The desired velocity is not a constant, but a variable which reduces towards zero as an agent gets close to the desired distance. This setting is more consistent with real-world scenario. The result of integrating $v_{ij}$ in varying desired velocity (Eq. 5.3) is compared with that of using $v_{ij}$ in a force component (Equation 5.2).**

In sum the idea of varying $v^0$ and $d^0$ is not conflicting with the control theory. The control targets are non-physics entities ($d_{ij}^0$ and $v_i^0$) and physics entities ($d_{ij}$ and $v_i$,) are updated to reach the targets. Also, the real situation is that physics entities ($d_{ij}$ and $v_i$,) are feedback to the human perception so that $d_{ij}^0$ and $v_i^0$ are also adjusted in certain conditions. The interplay between the physics entities ($d_{ij}$ and $v_i$,) and non-physics entities ($d_{ij}^0$ and $v_i^0$ ) forms a close-loop process as illustrated in Figure 5.6.

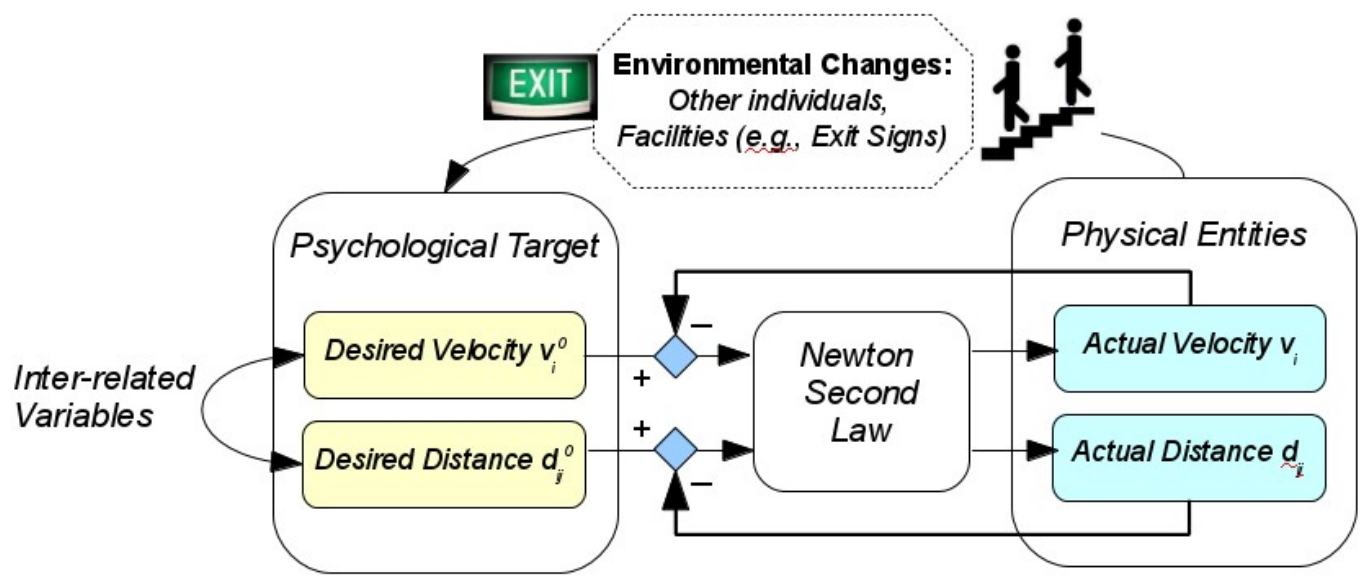

**Figure 5.6. Feedback Mechanism in Social Force Model: The idea of varying $d_{ij}^0$ and $v_i^0$ is also consistent with the control theory. The control targets are non-physics entities ($d_{ij}^0$ and $v_i^0$ ) and physics entities ($d_{ij}$ and $v_i$,) are updated to reach the targets. The interplay between the physics entities ( $d_{ij}$ and $v_i$,) and non-physics entities ($d_{ij}^0$ and $v_i^0$ ) forms a close-loop process.**

The focus of our study is modeling and understanding pedestrian from human perspective, and it relates to a new merging area called social statistical physics. Very importantly, the pedestrian dynamics is different from vehicle traffic or robot algorithms, and we critically consider human perception and cognition process as living bodies. Thus, our study provides a new angle to understand pedestrian behavior, and remarks are presented as below from the perspective of social science and psychology.

The oscillation may symbolize a kind of periodic stress. Recall Section 3.1 and 3.2 and stress is measured by difference between subjective opinion ($d_{ij}^0$ and $v_i^0$ ) and objective reality ($d_{ij}$ and $v_i$) as shown in Table 3.1, and the above analysis happens to indicate that stress is a periodic concept in a collective sense. Very interestingly, periodicity of stress in human behavior is also an existing subject in psychology, but it often refers to an individual-based study such as how personal mood or emotion changes with fluctuation of daylight, moon changes or seasons. The above computational models overcome an assumption that the individual is the crucial unit of cognition, but indicate that how such stress is possible to be generated from interaction of individuals ().

In sum the above analysis is meaningful to justify that stress exists in a collective sense, and it does not occur for a single individual if he or she has no contact with others. In our model stress is generated due to improper interpersonal distance or time-pressure, and it forms a kind of conscious wave that propagates in the crowd. So the entire crowd may become stressful, and behave within a certain frequency. This approach is balanced to cognitive science because the above models do not view cognition as a property of an individual mind, but as resulting from interactions among people and their environments [6].

The oscillation phenomenon is the origin of wave. Can we directly justify oscillation phenomenon of pedestrian crowd in reality? Sometimes it is possible for high-density crowd, but usually it is a sign of crowd tragedy like stampede. "It was like a huge wave of sea gushing down on the pilgrims" (P. K. Abdul Ghafour, Arab News), this is the crowd disaster in Mena on January 12, 2006. At a passageway when the crowd become more and more impatient, the conscious wave exiting in $d_{ij}^0$ and $v_i^0$ will become physical, and people thus start to push others in order to move forward. Such pushing force generated a pressure wave propagating in the crowd, and at a certain point, usually considered as a bottleneck, the wave may stop and it shows the power of destruction: someone may fall down and stampede happens there. In brief, as mentioned in Helbing et al., 2002, the social-force model may symbolizes various phenomena of human society at the macroscopic scale, and we think that such oscillation has a more profound meaning to explore in several aspects.

## VI. Stress and Fluid Dynamics of Pedestrian Crowd

In the previous sections we mainly discuss a kind of microscopic method, or as called many-particle simulation method, in which an individual's behavioral and psychological status is microscopically modeled by a set of mathematical equations or logic rules, and their interaction and collective behavior are calculated by numerical methods. This method is significantly boosted by advanced computing technology in recent years, and the merits exist in its capability of modeling a diversity of pedestrian characteristics, and we can observe the complex nonlinear dynamics in the simulation result. However, sometimes we lack analytical tools to deeply understand what we have observed in simulation. Thus, we also need the second method at the macroscopic level, which provides valuable intuition and predictions on the simulation results.

The second approach refers to an analytical model at the macroscopic level, aggregating many particle-like individuals into fluid-like model of crowd. In other words, the crowd fluid model (macro-level) is derived from homogenous individual equation (micro-level) based on physics laws and mathematical principles. The resulting model is a set of partial differential equations (PDE), which could be dated back to the well-known Eular Equation or Navior-Stocks Equations in 18th and 19th centuries. Since 1950s the equations were modified and applied in study of road traffic problem (Lighthill and Whitham, 1955), and further extended to pedestrian traffic problem (Al-nasur and Kachroo, 2006). The analytical solutions to these equations are often difficult in mathematics, but numerical solutions are available by advanced computing methods. Very importantly, analog of fluid dynamics is only valid to crowd at medium or high density, where continuity hypothesis holds for crowd flowing in a planar space. If there are only sparse individuals, continuity hypothesis cannot hold, and fluid-based analysis is not suitable for such low-density crowd. The resulting fluid model provides a practical perspective to explain crowd behavior at bottlenecks (e.g., narrow passages), where crowd density are sufficiently large and short-range physical interaction among people plays an important role. Such interactions are among the major cause of crowd disastrous events like stampede. The general framework of the fluid-based analysis is illustrated as below.

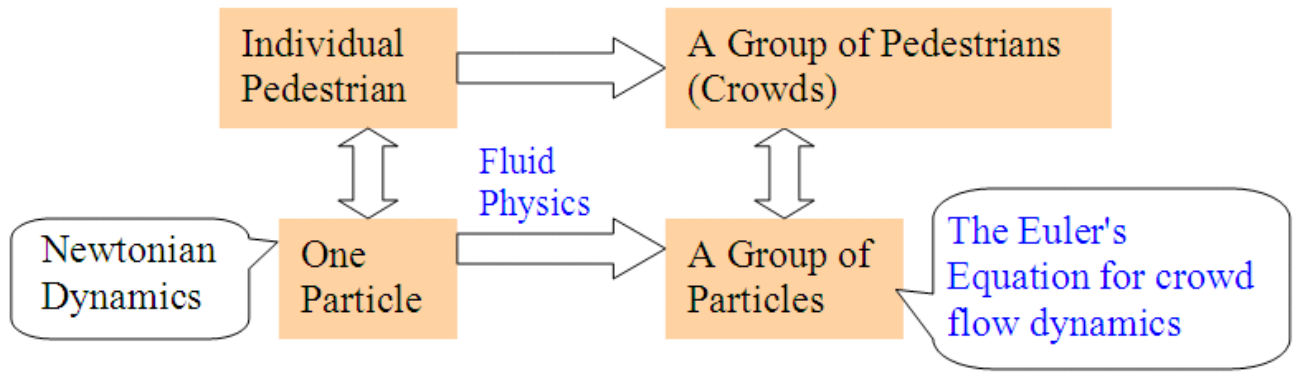

**Figure 6.1 Fluid-Based Analysis of Crowd Movement: The method is generally from fluid physics, which studies how individual agent interacts and emerge certain characteristics at the macroscopic level such as crowd speed and density.**

Although the methods could be either microscopic or macroscopic, they are different paths to the same problem, exploring the same phenomenon of crowd from different perspectives. Thus, advances in micro-models also facilitate model at macro-level, and verse visa. Based on this logic, this section is partly inspired from the microscopic model such as extended social force model in previous sections, and we introduce psychological variables from micro-simulation to fluid-based analysis of high-density crowd at bottlenecks (e.g, narrow doorways). The concepts of desired velocity, desired distance and stress measures are abstracted to the macroscopic level correspondingly. A set of partial differential equation (PDE) are essentially presented to describe how psychological intention of people interacts with physical characteristics of crowd motion (e.g., crowd speed and density), and it is an extension of Payne-Whitham model (Payne, 1971; Whitham, 1974, Helbing, 2001).

## 6.1. Crowd Fluid Dynamics in 2-Dimensional Space

As a crowd move on a planar surface, their movement can be considered as mass flowing with a specific rate in a two-dimensional space. The specific characteristics of crowd motion include:

(a) Flow density and mass: The flow density is the number of pedestrians per area unit, and it is defined by $\rho=(\mathrm{d}N)/(\mathrm{d}x\mathrm{d}y)$, where $\mathrm{d}N$ is the number of pedestrians in the area of $\mathrm{d}x\mathrm{d}y$ (Figure 6.2). The crowd density characterizes the distance of people. Let $m_0$ denote the average individual mass in the crowd, and the mass of the flow in area of $\mathrm{d}x\mathrm{d}y$ is $m=m_0\rho\mathrm{d}x\mathrm{d}y$.

(b) Interactions: As people move collectively, their physical interactions of people are characterized by surface pressure $P$**.** It is a two-dimensional analog of common pressure concept in three-dimensional space, and it is the lateral force per unit length applied on a line perpendicular to the force, namely $P=F/$( as shown in Figure 6.2(b). (Discontinuity of $P$ in the crowd fluid)

(c) Physical motion: The velocity of the moving crowd is denoted by $\boldsymbol{v}$, and it can be decomposed orthogonally as $v_x$ and $y_y$, characterizing the moving speed along $x$ axis and $y$ axis, respectively (See Figure 6.2).

(d) Motive Force: Human motion is self-driven. In physics, the motive force is commonly considered as frictions or pushing forces that people implement on the ground by feet. What differs creatures like human from non-creatures is that the motive force is intentionally generated by creatures such that they can decide where to go and how fast they move. Thus, the motive force is not only a physical concept, but also represents intentions of people in their mind. As a result, this paper presents the motive force by $\boldsymbol{f}^{self}=m\boldsymbol{a}^{self}$, where $\boldsymbol{a}^{self}$ is called self-acceleration and it indicates intentions of people. In a similar way, $\boldsymbol{a}^{self}$ can be decomposed as $a_x^{self}$ and $a_y^{self}$.

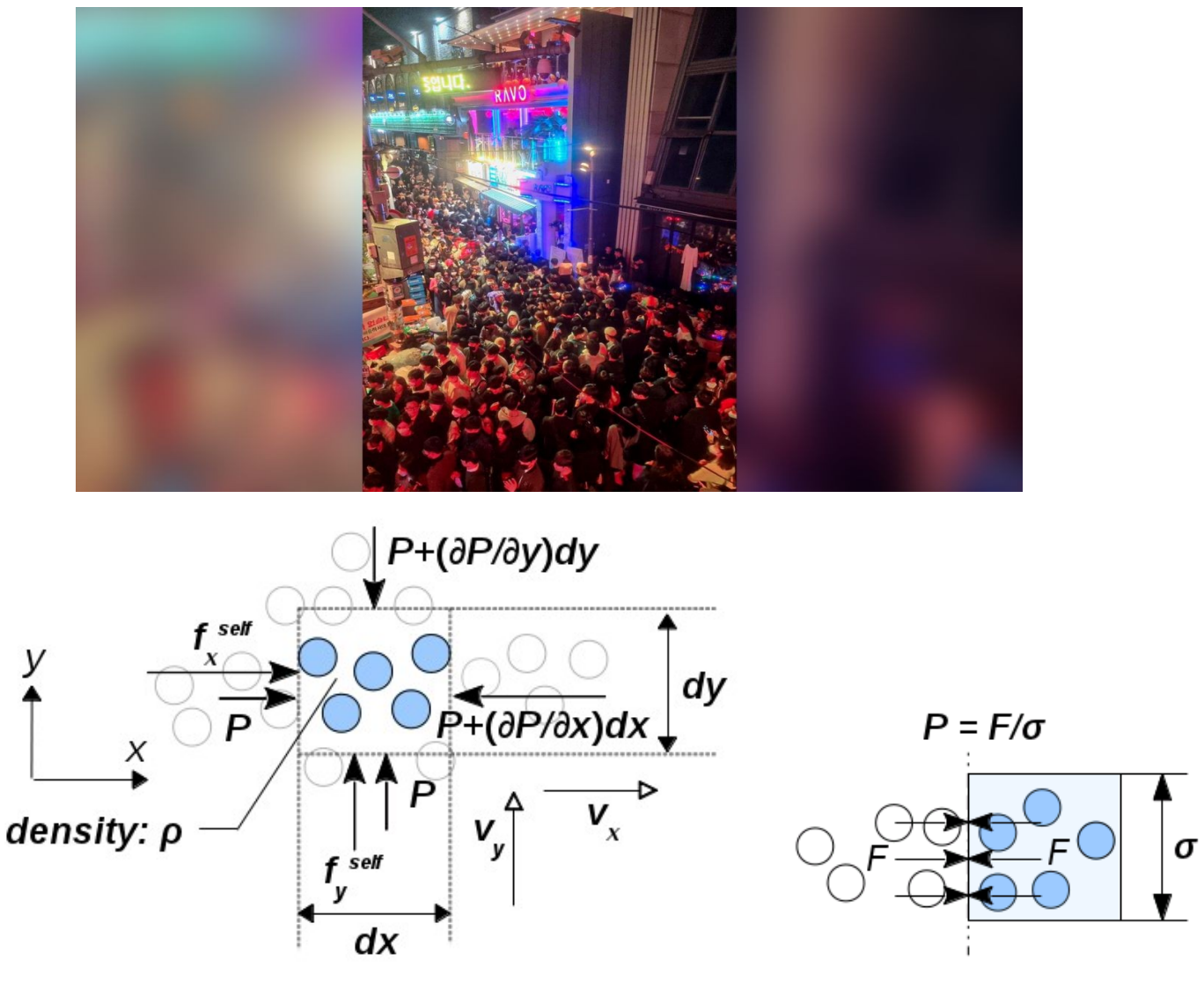

(a) (b)

**Figure 6.2 Crowd movement along a passageway: As a crowd move on a planar surface, their movement can be considered as mass flowing with a specific rate in a two-dimensional space. The flow characteristics such as crowd speed and density are all functions of position (*x,y*) and time *t*, and Newton Laws can also be applied in such analysis.**

Next, we will present crowd fluid equation in 2D space in a general sense. If readers are not familiar with fluid mechanics or not interested in knowing the mathematical description in 2D space, please move to subsection 6.2 on the next page to read the simplified analysis in 1D space.

The flow characteristics as presented above (i.e., $\rho$, $P$, $v$, $a^{self}$) are all functions of position $(x,y)$ and time $t$, and the crowd flow is unsteady flow in our analysis. With the assumption of the conservation of flow mass (i.e., mass continuity equation), we now study the moving crowd in the flow section of d$x$d$y$, where the mass is $m=m_0\rho \mathrm{d}x\mathrm{d}y$ . By the Newton Second Law we have

$$m_0\rho\left(\frac{\partial \boldsymbol{v}}{\partial t}+(\boldsymbol{v}\cdot\nabla)\boldsymbol{v}\right)=\boldsymbol{a}^{self}m_0\rho-\nabla \boldsymbol{P} \tag{6.1}$$

Equation (6.1) corresponds to Euler's Equation in fluid mechanics and it demonstrates the Newton Second Law ( $\Sigma F=ma$) for the crowd flow. Here we note that the derivation steps as above do not require any mathematical form of $\boldsymbol{a}^{self}$. Learning from the self-driving force and extended social-force as presented before, this paper specifies the subjective targets in people's mind by desired velocity $\boldsymbol{v}^d$ and desired density $\rho^d$, and the self-acceleration is given in a feedback manner as below, and $\boldsymbol{a}_v^{self}$ and $\boldsymbol{a}_\rho^{self}$ corresponds to the self-driving force and the extended social-force in the pedestrian model in Section 3 and 4.

$$\begin{gathered}\boldsymbol{a}^{self}=(\boldsymbol{a}_{\boldsymbol{v}}^{self}+\boldsymbol{a}_{\boldsymbol{\rho}}^{self})\\ \boldsymbol{a}_{\boldsymbol{v}}^{self}=k_1(\boldsymbol{v}^{\boldsymbol{d}}-\boldsymbol{v})=k_1[(v_x^d-v_x)\boldsymbol{i}+(v_y^d-v_y)\boldsymbol{j}]\\ \boldsymbol{a}_{\boldsymbol{\rho}}^{self}=k_2(\rho^d-\rho)\exp\left(\frac{\rho^d-\rho}{k_3}\right)\nabla\rho=k_2(\rho^d-\rho)\exp\left(\frac{\rho^d-\rho}{k_3}\right)\left(\frac{\partial\rho}{\partial x}\boldsymbol{i}+\frac{\partial\rho}{\partial y}\boldsymbol{j}\right)\end{gathered} \tag{6.2}$$

where $k_1$ $k_2$ and $k_3$ are parameters that weigh differently on targets of $v^d$ and $\rho^d$, and they have specific units to form the acceleration. Besides, the desired density $\rho^d$ and actual density $\rho$ correspond to the desired distance $d^0$ and actual distance $d$ in Equation (4.4).

The desired speed $v^d$ and desired density $\rho^d$ represent the psychological target in people's mind, specifying the speed and interpersonal distance that people desire to realize. The physical speed $v$ and density $\rho$ are the physical entities in the reality. As a result, the difference $v^d$ -$v$ and $\rho^d$ - $\rho$ show the gap between the human subjective wish and realistic situation, and they form the motive force to make the physical variables approaching towards the psychological targets.

According to Equation (4.4), if $d_{ij}$ is close to $d_{ij}^0$ , the extended social force is approximated in a linear form as below.

$$\boldsymbol{f}_{\boldsymbol{ij}}^{\boldsymbol{soc}} \approx \frac{A_{ij}}{B_{ij}}(d_{ij}^0 - d_{ij})\,\boldsymbol{n_{ij}} \quad \text{If} \quad d_{ij} \to d_{ij}^0 \tag{6.3}$$

Correspondingly, the motive force regarding the desired density can also be approximated in a linear manner when $\rho^d$ is close to $\rho$. As a result, Equation (6.2) is rewritten as below. In general, assumption of fluid only holds for high-density crowd (Helbing, et al, 2002). In other words, when people get sufficiently close with each other, they can be treated in analog with fluid and fluid analysis can be applied.

$$\begin{gathered}
\boldsymbol{a}^{\boldsymbol{self}} = (\boldsymbol{a}_{\boldsymbol{v}}^{\boldsymbol{self}} + \boldsymbol{a}_{\boldsymbol{\rho}}^{\boldsymbol{self}}) \\
\boldsymbol{a}_{\boldsymbol{v}}^{\boldsymbol{self}} = k_1(\boldsymbol{v}^{\boldsymbol{d}} - \boldsymbol{v}) = k_1[(v_x^d - v_x)\boldsymbol{i} + (v_y^d - v_y)\boldsymbol{j}] \\
\boldsymbol{a}_{\boldsymbol{\rho}}^{\boldsymbol{self}} = k_2(\rho^d - \rho)\nabla\rho = k_2(\rho^d - \rho)\left(\frac{\partial \rho}{\partial x}\boldsymbol{i} + \frac{\partial \rho}{\partial y}\boldsymbol{j}\right)
\end{gathered} \tag{6.4}$$

Equation (6.1) and (6.4) should be jointly used with other equations (e.g., the conservation of mass; boundary conditions) in order to fully describe the flow characteristics in a given geometric setting. By plugging Equation (6.4) into Equation (6.1) and using orthogonal decomposition. Equation (6.1) and (6.4) can be rewritten as below with the equation of mass continuity.

$$\begin{gathered}
m_0\rho\left(\frac{\partial v_x}{\partial t} + v_x\frac{\partial v_x}{\partial x} + v_y\frac{\partial v_x}{\partial y}\right) = \left(k_1(v_x^d - v_x) + k_2(\rho^d - \rho)\frac{\partial \rho}{\partial x}\right)m_0\rho - \frac{\partial P}{\partial x} \\
m_0\rho\left(\frac{\partial v_y}{\partial t} + v_x\frac{\partial v_y}{\partial x} + v_y\frac{\partial v_y}{\partial y}\right) = \left(k_1(v_y^d - v_y) + k_2(\rho^d - \rho)\frac{\partial \rho}{\partial y}\right)m_0\rho - \frac{\partial P}{\partial y} \\
\frac{\partial \rho}{\partial t} + \frac{\partial(\rho v_x)}{\partial x} + \frac{\partial(\rho v_y)}{\partial y} = 0
\end{gathered} \tag{6.5}$$

From the perspective of psychology studies, the gap between the psychological desire (i.e., $v^d$ and $\rho^d$) and the physical reality (i.e., $v$ and $\rho$) relates to how much stress people will be experiencing (Staal, 2004). In other words, the difference $v^d$-$v$ and $\rho^d$ - $\rho$ relates to the psychological concept of stress, and Equation (6.1) and (6.4) shows that accumulation of such stress will motivate certain behavior of people. In particular, the gap of speeds characterizes the time-related stress, which relates to the time-pressure in psychological research. The gap of density reflects the stress about the interpersonal relationship, and it focuses on social characteristics in the crowd. In particular, Equation (6.5) implies a feedback mechanism by which people's mind functions like a controller of their behavior: the targets in mind (i.e., $v^d$ and $\rho^d$) guide people to change their physical characteristics (i.e., $v$ and $\rho$), and such changes in the physical world are also feedback to people's mind. As a result, mind and reality interact in a closed-loop such that a balance can be reached between the psychological world of human mind and physical world of reality. Control theory can help to explore how $v$ and $\rho$ are changed given $v^d$ and $\rho^d$, and cognition science focuses on how people perceive the reality and thus adjust $v^d$ and $\rho^d$ based on their perceptions.

How to determine $v^d$ and $\rho^d$ in practical computing is a challenging task. The direction of $v^d$ could be given by solving a Possion Equation, where each candidate destination of movement is a sink point, and the flow solver calculates the route as the crowd move to the sink (Korhonen, 2018; Korhonen and Hostikka, 2010). The computation result is a 2D velocity field that guides evacuees to the destination selected. The magnitudes of $v^d$ and $\rho^d$ are not easy to be given because they are inter-related issues as mentioned in Section 5. In general, there should be preset values for $v^d$ and $\rho^d$, and in the simulation loop $v^d$ is first modified by any environmental stimuli that causes time-pressure, and $\rho^d$ is next adjusted by the difference $v^d$-$v$. In other words, the time-related stress is the very initial cause of motion, and it is transformed into space-related stress for high-density crowd. We will elaborate this idea from an energy perspective in Section 7.

### 6.2. Fluid-Based Analysis of Pedestrians at Bottlenecks

Next we will discuss a scenario when people move through a passageway as shown in Figure 6.3. Here we assume that the width of the passageway is relatively small compared to the length of the crowd flow. The flow is thus assumed to be homogeneous in the direction perpendicular to the passageway direction (i.e., along $y$ axis in Figure 6.2(a)), and the flow characteristics vary only along the passageway direction (i.e., along $x$ axis in Figure 6.2(a)). In brief, this section focuses on the crowd movement in a passageway like in a one-dimensional space, and the possible physical forces from walls $f_w$ are assumed to be and equal in magnitude and opposite in direction (See Figure 6.3). As a result, it yields $\partial P/\partial y$=0, $\partial \rho/\partial y$=0, $\partial v_x/\partial y$=0, $v^d_y$=0 and $v_y$=0. In other words, $v=v_x$ , $v^d=v_x^{\,d}$ and $f^{self}=f_x^{\,self}$. As shown in Figure 6.3, we now study the moving crowd in the flow section of $Y$d$x$, where the flow mass is $m=m_0\rho\,dx$ . As people try to accelerate along the positive direction of $x$ axis, the resistance comes from the surrounding people, and thus d$F$ = $P(x+dx)Y-P(x)Y$ is resistant to crowd motion. The motive force is $\boldsymbol{f}^{self}=m\boldsymbol{a}^{self}=\boldsymbol{a}^{self}m_0\rho\,dx$. By Newton Second Law we have

$$m_0\rho\, Y\,\mathrm{d}x\frac{\mathrm{d}v}{\mathrm{d}t}=f^{self}-[P(x+\mathrm{d}x)Y-P(x)Y] \tag{6.6}$$

The motive force is given by $\boldsymbol{f}^{self}=m\boldsymbol{a}^{self}$, where $\boldsymbol{a}^{self}$ is called self-acceleration and it indicates intentions of people. In a similar way, $\boldsymbol{a}^{self}=a_x^{\,self}$ in one-dimensional space.

$$m_0\rho\, Y\,\mathrm{d}x\frac{\mathrm{d}v}{\mathrm{d}t}=a^{self}m_0\rho\, Y\,\mathrm{d}x-[P(x+\mathrm{d}x)Y-P(x)Y] \tag{6.7}$$

Because $\partial P/\partial x=[P(x+dx)-P(x)]/dx$, then it follows

$$m_0\rho\frac{\mathrm{d}v}{\mathrm{d}t}=a^{self}m_0\rho-\frac{\partial P}{\partial x} \tag{6.8}$$

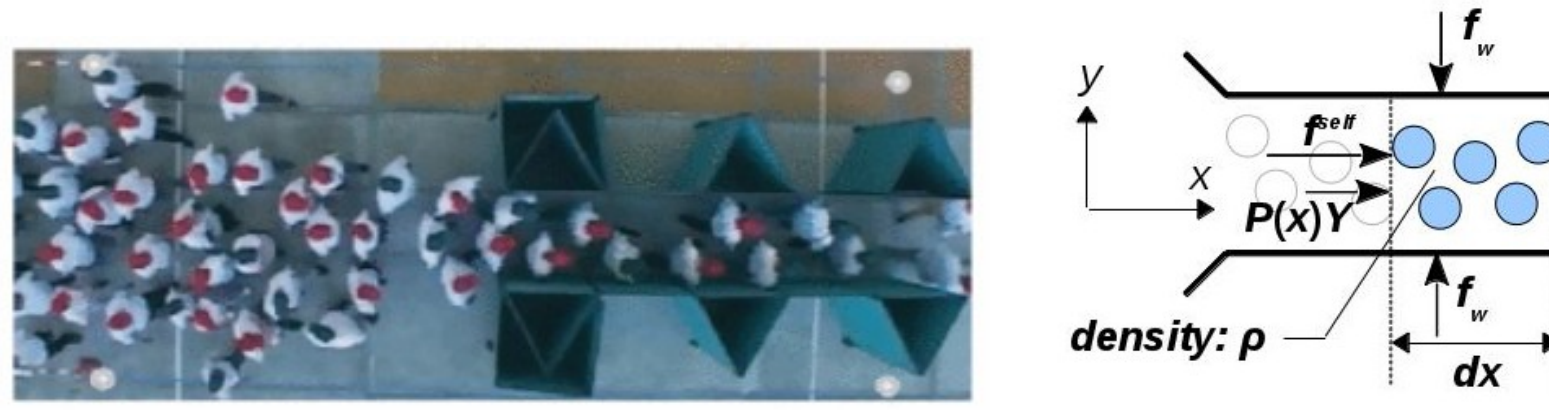

**Figure 6.3 Crowd movement along a passageway: Equation (6.5) is simplified for one-dimensional analysis, describing how crowd speed, density and pressure are formulated in Euler's Equation. (The real-world picture is cited from Hoogendoorn and Daamen, 2005)**

The above equation corresponds to Euler's Equation in fluid mechanics and it demonstrates the Newton Second Law ($ma=\Sigma F$) for crowd flow. Also, in one-dimensional space the flow characteristics as presented above (i.e., $\rho$, $P$, $v$, $a^{self}$) are all functions of position $x$ and time $t$, and it follows

$$m_0\rho\frac{1}{\mathrm{d}t}\left(\frac{\partial v}{\partial t}\mathrm{d}t+\frac{\partial v}{\partial x}\mathrm{d}x\right)=a^{self}m_0\rho-\frac{\partial P}{\partial x} \tag{6.9}$$

In fluid mechanics dv/dt is the Lagrangian derivative (material derivative) – the derivative following moving parcels in the fluid, and ∂v/∂t is the Eulerian derivative, which is the derivative of flow speed with respect to a fixed position.

$$m_0\rho\left(\frac{\partial v}{\partial t}+\frac{\partial v}{\partial x}\frac{\mathrm{d}x}{\mathrm{d}t}\right)=a^{self}m_0\rho-\frac{\partial P}{\partial x} \tag{6.10}$$

Note that $v$=d$x$/d$t$, it gives

$$m_0\rho\left(\frac{\partial v}{\partial t}+v\frac{\partial v}{\partial x}\right)=a^{self}m_0\rho-\frac{\partial P}{\partial x} \tag{6.11}$$

The above derivation steps do not specify any mathematical form of $\boldsymbol{a}^{self}$. As aforementioned, $\boldsymbol{a}^{self}$ should reflect the cognition process by which people perceive the physical world and form their motives. Based on Equation (6.4) the self-acceleration is given in a feedback manner as

$$a^{self}=k_1(v^d-v)+k_2(\rho^d-\rho)\frac{\partial\rho}{\partial x} \tag{6.12}$$

where $k_1$ $k_2$ are parameters that weigh differently on targets of $v^d$ and $\rho^d$, and they have specific units to form the acceleration. As aforementioned, the difference $v^d$-$v$ and $\rho^d$-$\rho$ show the gap between the human subjective wish and realistic situation, and they form the motive force to make the physical entities approaching towards the psychological targets. From the perspective of psychology, the gap between the psychological desire (i.e., $v^d$ and $\rho^d$) and the physical reality (i.e., $v$ and $\rho$) relates to how much stress people are experiencing (Staal, 2004). The gap of speeds $v^d$-$v$ characterizes the time-related stress, which relates to time-pressure in psychological research, and it corresponds to the self-driving force in Helbing, Farkas, and Vicsek, 2000.

The gap of density $\rho^d - \rho$ reflects the stress about the interpersonal relationship, and it corresponds to the extended social force which focuses on social characteristics in the crowd.

$$m_0\rho\left(\frac{\partial v}{\partial t}+v\frac{\partial v}{\partial x}\right)=\left(k_1(v^d-v)+k_2(\rho^d-\rho)\frac{\partial \rho}{\partial x}\right)m_0\rho-\frac{\partial P}{\partial x}$$

$$\frac{\partial \rho}{\partial t}+\frac{\partial(\rho v)}{\partial x}=0 \tag{6.13}$$

Mathematically, Equation (6.13) are to be jointly used with boundary conditions in order to fully describe the flow characteristics in a given geometric setting. Based on the existing theory in partial differential equations, the solution should be a wave function, which implies that the gap of $v^d$ -$v$ and $\rho^d - \rho$ are periodic functions. In other words, $v$ and $\rho$ will sway around the target $v^d$ and $\rho^d$, and finally converge to the target value $v^d$ and $\rho^d$. If $v^d$ and $\rho^d$ are dynamically changed with time, then $v$ and $\rho$ will also track such changes with a time delay. Next, we will brief discuss the solution of Equation (6.13). If there is no physical interactions among people, then $\partial P/\partial x=0$, and it yields

$$\left(\frac{\partial v}{\partial t}+v\frac{\partial v}{\partial x}\right)=\left(k_1(v^d-v)+k_2(\rho^d-\rho)\frac{\partial \rho}{\partial x}\right)$$

$$\frac{\partial \rho}{\partial t}+\frac{\partial(\rho v)}{\partial x}=0 \tag{6.14}$$

$$\frac{\partial v}{\partial t}+v\frac{\partial v}{\partial x}-k_2(\rho^d-\rho)\frac{\partial \rho}{\partial x}=k_1(v^d-v)$$

$$\frac{\partial \rho}{\partial t}+\rho\frac{\partial v}{\partial x}+v\frac{\partial \rho}{\partial x}=0 \tag{6.15}$$

The above equation is a hyperbolic partial differential equation. As below is the typical approach for analysis of the equation.

$$\frac{\partial v}{\partial t}=-v\frac{\partial v}{\partial x}+k_2(\rho^d-\rho)\frac{\partial \rho}{\partial x}+k_1(v^d-v)$$

$$\frac{\partial \rho}{\partial t}=-\rho\frac{\partial v}{\partial x}-v\frac{\partial \rho}{\partial x} \tag{6.16}$$

$$\begin{pmatrix}\dfrac{\partial v}{\partial t}\\[2mm] \dfrac{\partial \rho}{\partial t}\end{pmatrix}=\begin{pmatrix}-v & -k_2(\rho-\rho^d)\\ -\rho & -v\end{pmatrix}\begin{pmatrix}\dfrac{\partial v}{\partial x}\\[2mm] \dfrac{\partial \rho}{\partial x}\end{pmatrix}+\begin{pmatrix}k_1(v^d-v)\\ 0\end{pmatrix} \tag{6.17}$$

Here we take the state variable as $U=(v,\ \rho)^T$, and the Jacobian matrix is

$$A(U)=\begin{pmatrix}v & k_2(\rho-\rho^d)\\ \rho & v\end{pmatrix} \qquad |A(U)-I\lambda|=\begin{vmatrix}v-\lambda & k_2(\rho-\rho^d)\\ \rho & v-\lambda\end{vmatrix} \tag{6.18}$$

Then the eigenvalue are calculated as

$$\lambda=v\pm\sqrt{k_2\rho(\rho-\rho^d)} \tag{6.19}$$

The above result shows a hyperbolic partial differential equation if $\rho>\rho^d$ . Otherwise the equation is elliptical or parabolic. This result shows isotropic nature of the model if the crowd density is sufficiently large. For low-density crowd the information spreads anisotropically. This is reasonable because pedestrians normally perceive ones in front, but not the ones behind, and thus the information cannot spread isotropically. This is why we add an anisotropic factor $0 < \lambda_i < 1$ in Equation (2.3), (3.1) and (4.4). However, pedestrian are able to feel the ones standing behind only if they are very close to each other, which means the crowd density is sufficiently large. In this situation the information does spread from behind, and the ones in behind could inform the people in front to move faster (See Figure 6.2), or physically push them to compress the crowd.

This is usually the cause of crowd disasters like stampede. The numerical method of the above pedestrian dynamics in 1D space is formulated as below.

$$\frac{v_i(t+1)-v_i(t)}{T}=-v_i(t)\frac{v_i(t)-v_{i-1}(t)}{\Delta}+k_2(\rho_i^d(t)-\rho_i(t))\frac{\rho_i(t)-\rho_{i-1}(t)}{\Delta}+k_1(v_i^d(t)-v_i(t))$$

$$\frac{\rho_i(t+1)-\rho_i(t)}{T}=-\rho_i(t)\frac{v_i(t)-v_{i-1}(t)}{\Delta}-v_i(t)\frac{\rho_i(t)-\rho_{i-1}(t)}{\Delta} \quad (6.20)$$

where i=1, 2, 3 … N and t=0, 1, 3 … M.

$$v_i(t+1)=v_i(t)+\frac{T}{\Delta}\Big(-v_i(t)(v_i(t)-v_{i-1}(t))+k_2(\rho_i^d(t)-\rho_i(t))(\rho_i(t)-\rho_{i-1}(t))\Big)+Tk_1(v_i^d(t)-v_i(t))$$

$$\rho_i(t+1)=\rho_i(t)+\frac{T}{\Delta}\Big(-\rho_i(t)(v_i(t)-v_{i-1}(t))-v_i(t)(\rho_i(t)-\rho_{i-1}(t))\Big) \quad (6.21)$$

Although fluid mechanics has long been established in physics, the analytical solution is often difficult. Most equations can only be solved by numerical methods, such as finite difference method (FDM) or finite volume method (FVM). One major problem is that such numerical methods requires much more computational time than many-particle simulation. However, the numerical solutions are too abstracted and fluid-like, and they seems far away from real-world scenarios. They may provide some valuable insights, but not similar to many-particle simulation as introduced in previous sections. Thus, we will not further discuss the numerical solution of PDE in this article.

Last but not least, we compare the traditional vehicular flow with the above pedestrian flow model. Note that vehicular flow is strictly confined by lane markings while pedestrians are not subject to any lane restrictions or queuing laws. Therefore, although shock waves and stop-and-go waves are widely discussed in the classical car following problem (e.g., as the traffic light turns from red to green), they are not common phenomena in pedestrians unless queuing behavior strictly keep the sequential order of their motion. In disorderly crowd, competitive behavior may exist and interrupt the stop-and-go wave. For example, someone under much time-pressure may try to move faster than surrounding others and and take over them in the flow. Therefore, shock waves and stop-and-go waves may partly exist for pedestrian flows, but they are not obvious in real-world observations because of non-queuing behavior in pedestrian crowd. As for analytical discussion on shock waves and stop-and-go waves, please refer to Whitham, 1974 for more detailed discussion.

## VII. Energy-Based Analysis of Pedestrian Crowd

This section will continue with one-dimensional analysis of crowd flow, and introduce a Bernoulli-like approach to explain how psychological intention of people interacts with physical characteristics of crowd motion (e.g., crowd speed and density). By aggregating the Newtonian motion of individual pedestrians to the flow of crowds, an energy balance equation is derived in this section to explore how energy is transformed from the psychological world of human mind to the physical world of universe. Such energy-based analysis helps to bridge a gap among psychological findings, pedestrian models and simulation results, and it further provides a new perspective to understand the paradoxical relationship of subjective wishes of human and objective result of their deeds in reality.

### 7.1 Energy-Based Analysis of Pedestrian Crowd

When the crowd flow converges into the steady state, it implies ∂v/∂t=0 and the Eulerian derivative becomes zero. For such steady flow, the flow characteristics (i.e., $\rho$, $P$, $v$, $a^{self}$) are time-invariant, and energy-based analysis is commonly applied. By taking the dot product with d$s$ – the element of moving distance – on both sides of Equation (6.11), an energy equation can be derived, which corresponds to the well-known Bernoulli equation in fluid mechanics. In particular, the element of distance in one-dimensional space is d$s$=d$x$, and thus we have

$$m_0\rho v\frac{\partial v}{\partial x}\mathrm{d}x=a^{self}m_0\rho\,\mathrm{d}x-\frac{\partial P}{\partial x}\mathrm{d}x \quad (7.1)$$

Because the flow characteristics (i.e., $\rho$, $F$, $v$, $a^{self}$) are only functions of position $x$ for steady flow, it gives

$$m_0 v\,\mathrm{d}v=a^{self}m_0\,\mathrm{d}x-\frac{\mathrm{d}P}{\rho} \quad (7.2)$$

$$m_0 \,\mathrm{d}\left(\frac{v^2}{2}\right)+\frac{\mathrm{d}P}{\rho}=a^{self}\,m_0\,\mathrm{d}x \tag{7.3}$$

Since $m_0$ is the average individual mass and does not depend on moving speed $v$, the above equation can be integrated. The physical interactions are repulsive among people, and $P \geqq 0$. An energy balance equation is obtained as below

$$\frac{m_0 v^2}{2}+\int_0^P \frac{\mathrm{d}P}{\rho}=\int_{x_0}^{x} m_0 a^{self}\,\mathrm{d}x+C \tag{7.5}$$

Based on Equation (6.12) it further gives

$$\frac{m_0 v^2}{2}+\int_0^P \frac{\mathrm{d}P}{\rho}=\int_{x_0}^{x} m_0 k_1 (v^d-v)\,\mathrm{d}x+\int_{x_0}^{x} m_0 k_2 (\rho^d-\rho)\frac{\partial \rho}{\partial x}\mathrm{d}x+C \tag{7.6}$$

Similar to the well-known Bernoulli Equation in fluid mechanics, Equation (7.6) can be interpreted by the principle of energy conservation, where the psychological drive can be considered as a special form of potential energy that arises from crowd opinion, and its behavioral manifestations are energy in physical forms.

The left side of Equation (7.6) includes energy in physical forms – kinetic energy and static energy. In addition, if the crowd movement is not horizontal, but on a slope, the gravity should be taken into account by including the gravitational potential $m_0gh$ on the left side of Equation (7.6), where $h$ is the altitude of the crowd position and $g$ is the gravitational acceleration. The kinetic energy is the common form that describes the energy regarding crowd motion. The static energy is an integral form that characterizes the physical interactions of people. The physical interaction comes into existence when the crowd density exceeds a certain limit. Therefore, $P$ can also be considered as a function of crowd density $\rho$ A mathematical expression of the static energy is exemplified as below. Let $\rho_0$ represent the crowd density when people start to have physical contact, and the interaction force is given by

$$P=K\,\eta\left(\rho-\rho_0\right) \qquad\qquad \int_0^P \frac{\mathrm{d}P}{\rho}=K\,\eta\left(\ln\rho-\ln\rho_0\right) \tag{7.7}$$

where $K$ is a positive parameter. $\eta(\cdot)$ is a piecewise function such that $\eta(x)=0$ if $x<0$ and $\eta(x)=x$ if $x>=0$. The interaction force thus becomes nonzero when $\rho>\rho_0$. The resulting static energy is given by Equation (7.7). In general, the moving crowd is a compressible flow: the crowd density varies in different places of the flow. Thus, the static energy is a function of density $\rho$.

The right side of Equation (7.6) is energy involving crowd opinion, and it is called motivation energy or stress energy in this paper. This special kind of energy is expressed by an integral term, and it follows the typical formula that is the integral product of a force and moving distance along the direction of the force. The motivation energy in Equation (7.6) consists of two terms: one term drives people to adjust speed in a temporal space and the other one motivates people to adjust their social distance with others. This corresponds to the driving force and social force in the social force model. An interesting topic is that Equation (7.6) extends the law of energy conservation from the physical world of universe to the psychological world of human mind, implying potential transformation of motivation energy into certain physical form. In fact, the energy-based equation shows that energy arises in mind when people have desire doing something, and it will be ultimately transformed into certain physical energy in reality. In other words, energy in mind cannot vanish by itself, but must find an outlet to the physical world.

Table 7.1 Conservation of Energy in Crowd Collective Motion

| | | |
|---|---|---|
| Kinetic Energy | $\frac{m_0 v^2}{2}$ | Energy in Physical World |
| Static Energy | $K\,\eta\left(\ln\rho-\ln\rho_0\right)$ | ↑↓ |
| Motivation (Psychological Features) | $\int_{x_0}^{x} m_0 a^{self}\,\mathrm{d}x$ | Energy from Conscious Mind |

Table 7.2 Energy Balance in Crowd Movement

| Kinetic Energy | ⇦ | Static Energy |
| --- | --- | --- |
| $0.5\, m_0 v^2$ | ⇨ | $K\,\eta\left(\ln\rho - \ln\rho_0\right)$ |
| Influence ⇧ | | ⇩ Feedback |
| $\int_{x_0}^{x} m_0 k_1 (v^d - v)\,\mathrm{d}x$ | ⇦ | $\int_{x_0}^{x} m_0 k_2 (\rho^d - \rho)\frac{\partial\rho}{\partial x}\,\mathrm{d}x$ |
| Self-Driving Characteristic | ⇨ | Social Characteristic |

The motivation energy characterizes the collective opinion of people in mind. From the perspective of psychological principles, the variables of $v^d$ and $\rho^d$ thus have much freedom because they exist in people's mind. However, as people realize $v^d$ and $\rho^d$, their behavior are not free any more since certain realistic factors confine their deed in the physical world (e.g, the size of a passage may not permit all the people to move as fast as desired). If the physical variables reach the maximum while people still desire increasing them, the motivation energy will not be transformed to the physical forms as desired. In this situation, the stronger is the subjective wish in people's mind, the worse may become the situation in the reality, showing a paradoxical relationship of subjective wishes of human and objective result in reality. An example in this kind is the “faster-is-slower” effect as shown in Helbing, Farkas, and Vicsek, 2000. Just as the proverb says – “More haste, less speed.”

## 7.2 Doorway Scenario and Egress Performance

The energy-based analysis provides us a new perspective to reinterpret the faster-is-slower effect and Yerkes–Dodson law. If the motivation energy is transformed properly so that people are able to speed up, the psychological drive of motion will accelerate the crowd, and faster-is-faster effect shows up. In contrast, when the motivation energy cannot be transformed to the physical forms as desired, the faster-is-slower effect comes to being. Very importantly, the passage capacity determines the maximal amount of motivation energy that can be transformed to the physical forms, and it determines a critical threshold: below the threshold the psychological drive as expressed by the motivation energy is transformed to the kinetic energy and the crowd can accelerate as desired. Above the threshold the kinetic energy reaches the maximum, and excess of psychological drive will be transformed to the static form, resulting in an increase of crowd density. This answers the question about when the motivation energy is transformed to the kinetic form, and when to the static form. To describe this scenario in consistency with real-life situation, Equation(7.6) is further reshaped as below.

$$\frac{m_0 v^2}{2} + \int_0^{\rho} \frac{\mathrm{d}\left(K\,\eta(\rho-\rho_0)\right)}{\rho} = \int_{x_0}^{x} m_0 k_1 (v^d - v)\,\mathrm{d}x + \int_0^{\rho} m_0 k_2 (\rho^d - \rho)\,\mathrm{d}\rho + C \tag{7.8}$$

$$\frac{m_0 \rho v^2}{2} + \int_0^{\rho} \mathrm{d}\left(K\,\eta(\rho-\rho_0)\right) - \int_0^{\rho} m_0 k_2 (\rho^d - \rho)\,\rho\,\mathrm{d}\rho = \int_{x_0}^{x} m_0 k_1 (\rho v^d - \rho v)\,\mathrm{d}x + C \tag{7.9}$$

$$\frac{m_0 \rho v^2}{2} + K\,\eta(\ln\rho - \ln\rho_0) - \int_0^{\rho} m_0 k_2 (\rho^d - \rho)\,\rho\,\mathrm{d}\rho = \int_{x_0}^{x} m_0 k_1 (q^d - q)\,\mathrm{d}x + C \tag{8.0}$$

In Equation(7.9) physical crowd flow is measured by $q=\rho v$, which reflects the physical motion that the crowd is able to realize, and it indicates the number of pedestrians who are physically able to pass through a passage per time unit. In contrast, the desired flow is indicated by $q^d=\rho v^d$, which describes the desire of movement in a psychological sense. The difference of $q^d–q$ is an indicator of stress in term of flow dynamics.

Furthermore, the capacity of a passage is $c=\max\{q\}$, the maximal number of people who can pass through the passage per time unit. With the above flow concepts, the disorder and blocking effect is modeled as: when the flow demand $q^d=\rho v^d$ is below the passage capacity, crowd can move as fast as desired, and $q$ is equal to $q^d$. If the flow demand $q^d$ exceeds the limited capacity, the probability of disorder and blocking will increases with the increase of the overage, resulting in a decrease of the physical flow rate in a nonlinear fashion in Figure 7.1.

In a psychological sense, the difference $q^d–q$ also indicates a kind of time-related stress which motivates people to move fast. At bottlenecks (e.g., a narrow staircase) people may not pass through fast due to the limited passage capacity, and the time-related stress is transformed into space-related interpersonal stress, resulting increase of desired density $\rho^d$. The desired density $\rho^d$ keeps increasing to raise physical density ρ in waiting pedestrians. If the physical density $\rho$ exceeds the critical value $\rho_0$ in (7.7), physical interactions occur and the risk of disorder and stampede increases in pushy crowd. As a consequence, the pedestrian flow will decrease. Further increase of the crowd physical density $\rho$ could result in disorder events (e.g., jamming or stampeding). In this light our crowd flow model and energy-based analysis also reiterates the Yerkes–Dodson law: moderate stress improves the performance (i.e., speeding up crowd flow) while excessive stress impairs it (e.g., disordering and jamming). The relationship of the desired flow $q^d=\rho v^d$ and physical flow $q=\rho v$ is plotted as shown in Figure 7.1, where $q^d=\rho v^d$ indicates the collective demand of crowd movement and $q=\rho v$ indicates the physical motion that the crowd realize. Corresponding to Figure 3.1 this figure reiterates the Yerkes-Dodson law at the macroscopic level, and $q^d=\rho v^d$ is the stress indicator, which represents the motivation level at the macroscopic level.

Very importantly, we note that the traditional vehicular flow problem assumes that the desired velocity $v^d$ reduces to zero at some maximum density as in traffic jams, e.g., the Greenshields model $v^d=v_f(1-\rho/\rho_{jam})$, Kladek model $v^d=v_f(1\text{-}\exp(1/\rho-1/\rho_{jam}))$, where $v_f$ is the freeway speed at density $\rho=0$, and $v^d$ reduces to zero at $\rho_{jam}$, which is density in traffic jams. These models were formulated for vehicular flow, where desired driving speed $v^d$ drops with increasing density $\rho$, and physical speed $v$ converges to desired speed $v^d$ to reach the steady states. However, by integrating desired density $\rho^d$ in the above fluid-based analysis, these empirical rules seems problematic, especially for understanding risk in disorderly crowd incidents. In fact, increasing pedestrian density initially exhibits a kind of collective intelligence to raise flow rate $q=\rho v$, but if $q=\rho v$ reaches the maximum, impatient pedestrians may behave competitively, and it seems problematic to assume that $v^d$ decreases to zero in jamming crowd, and it contradicts with the paradoxical understanding of the faster-is-slower effect.

Thereby, it is conceptually questionable to assume that pedestrians' desired moving speed is reduced to zero even if researchers do observe that their physical flow is turned out to be zero in physics measurement. The physical flow speed is reduced because the limited capacity of space cannot meet the collective demand by too many people, not because people desire standing still and not moving forward. If everyone is standing still and nobody become impatient, there will be no crowd collapse or stampede. Thus, reducing desired speed produces a simulation that looks plausible, but it is questionable in the fundamental concepts or principles. Thus, as suggested by Dr. Helbing and his co-workers, the empirical flow diagram of $q(\rho)=\rho v(\rho)$ is partly misleading in study of the fundamental cause of crowd surge incidents.

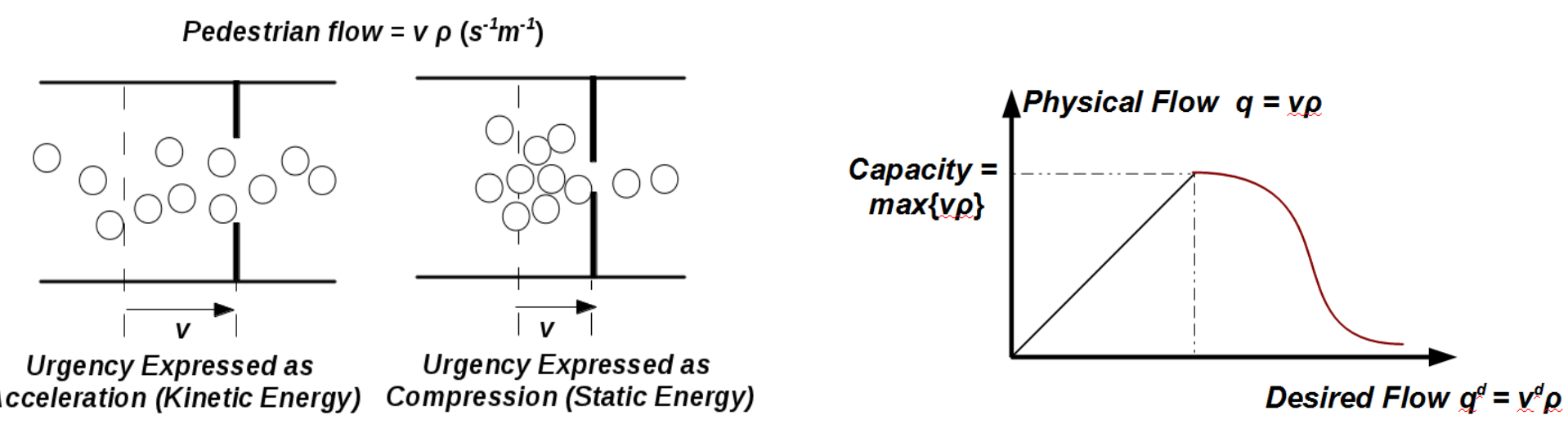

**Figure 7.1 Relationship of the desired flow $q^d=\rho{}^d$ and physical flow $q=\rho v$: $\rho{}^d$ indicates the collective demand of crowd movement and $\rho v$ indicates the physical motion that the crowd realize. Corresponding to Figure 3.1 the above figure also reiterates the Yerkes-Dodson law at the macroscopic level, and $q^d=\rho{}^d$ is the stress indicator, which indicates the motivation level at the macroscopic level, and moderate stress improves the crowd performance (i.e., speeding up crowd flow) while excessive stress impairs it (e.g., disordering and jamming).**

Continuing with the above energy-based analysis we will next take the gravitational force and fluctuation random force into account, and explain the risk of "down-slope" and "heating effect" in crowd egress. Down-slopes are commonly known as stairways, where the gravity is effective. As a result, the gravitational potential is necessarily taken into account in Equation (7.6), and it can either facilitate or impede acceleration or compression of a crowd. When people desire moving faster ($v^d>v$), the gravity facilitates acceleration on a down-slope while resists such acceleration on an up-slope. A similar effect is also for compression. Because people usually move downstairs in egress, acceleration and compression will be more intensive in a down-slope, and the risk of disorder and blocking is thus increased.

The heating effect was known for the "heating-by-freezing" in Helbing et al., 2002, and it implies that fluctuation force in Equation (2.1) increases on average (Helbing et al., 2002). Increasing fluctuation $\xi_i$ is not meaningful if there are only sparse individuals in the pedestrian crowd. Thus, in many-particle simulation as discussed before we do not consider this fluctuation force. However, for high-density crowd the things are much different, and increase of $\xi_i$ means the individuals gain more vigor or inner energy such that their stochastic motion become more and more intensive. Therefore, we will especially add the heating effect in the fluid dynamics of high-density crowd.

Table 7.3 Conservation of Energy with Fluctuation Effect and Gravitational Effect

| | | |
|---|---|---|
| Kinetic Energy | $\frac{m_0 v^2}{2}$ | Energy in Physical World ↑↓ Energy from Conscious Mind |
| Static Energy | $\int_0^P \frac{dP}{\rho}$ | |
| Gravitational Energy | $m_0 gh$ | |
| Motivation & Fluctuation (Psychological Features in Collective Opinions) | $\int_{x_0}^{x} m_0 a^{self}\,\mathrm{d}\,x+\int_{x_0}^{x} m_0 a^{\xi}\,\mathrm{d}\,x$ | |

From the perspective of energy-based analysis people often feel excited when watching sports games, attending concert or joining parties, and they usually exhibit more vigor and release more inner energy in these occasions. Even if they are not in physical motion, they will release such energy by shouting or other behavior. So stadium, concert or nightclubs are a kind of public places where people's inner energy will find an outlet to the physical world, and such effect on crowd is similar to heating process as mentioned in Helbing et al., 2002. Corresponding to the fluctuation force in Equation (2.1), a fluctuation acceleration is added in Equation (6.11), and it indicates the heating effect on the crowd. For heating effect the fluctuation force is not toward a certain direction and it represents irrational mind of people. In contrast the desired velocities and desired distance are deterministic in directions and represent rational mind of people.

$$\frac{m_0 v^2}{2}+\int_0^P \frac{dP}{\rho}+m_0 gh=\int_{x_0}^{x} m_0 a^{self}\,\mathrm{d}\,x+\int_{x_0}^{x} m_0 a^{\xi}\,\mathrm{d}\,x+C \qquad (8.1)$$

When the "heated" crowd are further compressed at a downhill bottleneck in egress, the risk of disorder and stampede is much higher than in other places. In a list of historical events of stampeding in Still, 2016, it suggests that certain gathering places such as stadium are more frequent to occur stampede in emergency egress. Similar evidence is also listed in Helbing et. al., 2002, and one of the crucial reasons is that bottlenecks in a stadium are often not horizontal, but on stairways (See Figure 7.2). Another important reason is that people usually feel excited when watching football games, and the crowd are thus "heated" as Helbing described. From the practical viewpoint, it is almost impossible to clam down the crowd who join a sport event in stadium, but it is feasible to design egress facilities such that the bottleneck (e.g, a narrow passageway) are not placed on down-slope areas. In other words, it is better to design the downstairs ways in a wide or open area, and when people get gathered at the bottleneck, the passageway should be horizontal.

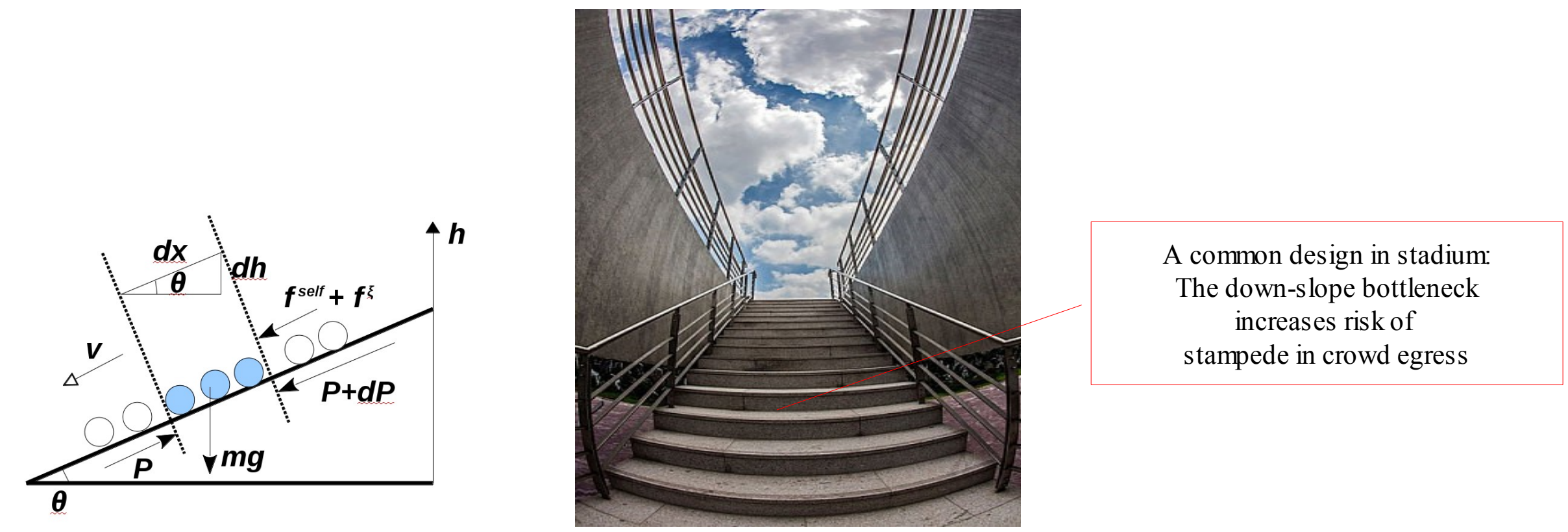

**Figure 7.2 A Bottleneck Stairway at Stadium: When the "heated" crowd are further compressed at a downstairs bottleneck in egress, the risk of disorder and stampede is much higher than in other places. In a list of historical events of stampeding in Still, 2016, it suggests that certain gathering places such as stadium are more frequent to occur stampede in emergency egress.**

## VIII. CONCLUSION

Human motion is self-driven and self-adapted to the environment, and the forces are generated by intention of people in a psychological sense. How to capture this characteristic in consistency with known physics laws is meaningful to understand crowd behaviors. By integrating psychological principles to Newtonian motion of pedestrian crowd, we reinterpret the well-known social-force model by using the psychological concept of stress, and this individual-based model is further aggregated into crowd fluid dynamics at the macroscopic level. The concept of stress at both micro-and-macro scales is interpreted by the difference between the desired entities in human mind (i.e., desired velocity, desired interpersonal distance, desired crowd density) and the physical entities in reality (i.e., actual velocity, actual interpersonal distance, actual crowd density), and the resulting crowd dynamics describes how environmental stressors (e.g., surrounding people, hazard and exit signs) influence collective behavior in both normal and emergency condition. In brief, our study at micro-and-macro level helps to bridge a gap among psychological findings, pedestrian models and simulation results, and it further provides a new perspective to understand the paradoxical relationship of subjective wishes of human and objective result of their deeds in reality.

## ACKNOWLEDGMENTS

The author is grateful to Peter B. Luh and Kerry L. Marsh for helpful comments on earlier work in University of Connecticut. The author is also thankful to Timo Korhonen and Mohcine Chraibi for helpful discussion in simulation work of FDS+Evac. The author is sincerely thankful to Xiaoda Wang, Prof. G. Keith Still, Prof. Dirk Helbing for helpful discussion. The author appreciates the research program funded by NSF Grant # CMMI-1000495 (NSF Program Name: Building Emergency Evacuation - Innovative Modeling and Optimization).

## SUPPLEMENTARY DATA

The supplementary data to this article are available online at https://github.com/godisreal/test-crowd-dynamics and https://github.com/godisreal/crowdEgress. If you have any comment or inquiry about the testing result, please feel free to contact me at wp2204@gmail.com or start an issue in the repository.